\documentclass[11pt,a4paper]{article} 
\pdfoutput=1 

\usepackage{jheppub}

\usepackage{amsmath,amssymb,multirow,graphicx}
\usepackage[numbers,sort&compress]{natbib}
\usepackage{subfigure}

\title{Neutrino observables from predictive flavour patterns} 

\author[a]{Lu\'{\i}s M. Cebola,}
\author[a]{David Emmanuel-Costa,}
\author[b,a]{and Ricardo Gonz\'{a}lez Felipe}

\emailAdd{luismcebola@tecnico.ulisboa.pt}
\emailAdd{david.costa@tecnico.ulisboa.pt}
\emailAdd{ricardo.felipe@tecnico.ulisboa.pt}

\affiliation[a]{Departamento de F\'{\i}sica and Centro de F\'{\i}sica Te\'{o}rica de Part\'{\i}culas - CFTP, Instituto Superior T\'{e}cnico, Universidade de Lisboa, Avenida Rovisco Pais, 1049-001 Lisboa, Portugal}
	
\affiliation[b]{ISEL - Instituto Superior de Engenharia de Lisboa, Instituto Polit\'ecnico de Lisboa,\\ Rua Conselheiro Em\'{\i}dio Navarro 1959-007 Lisboa, Portugal}

\abstract{
We look for predictive flavour patterns of the effective Majorana neutrino mass matrix that are compatible with current neutrino oscillation data. Our search is based on the assumption that the neutrino mass matrix contains equal elements and a minimal number of parameters, in the flavour basis where the charged lepton mass matrix is diagonal and real. Three unique patterns that can successfully explain neutrino observables at the $3\sigma$ confidence level with just three physical parameters are presented. Neutrino textures described by four and five parameters are also studied. The predictions for the lightest neutrino mass, the effective mass parameter in neutrinoless double beta decays and the CP-violating phases in the leptonic mixing are given.
}

\date{\today}

\begin{document}

\maketitle
\flushbottom

\section{Introduction}
\label{sec:intro}

Neutrino masses constitute a solid evidence of physics beyond the standard model. In the absence of a compelling theory to explain the origin of the lepton flavour structure, one expects the new physics to shed some light in the flavour puzzle. From the theoretical point of view, a natural approach is to restrict the number of free parameters in the lepton flavour sector so that the theory becomes more predictive. One possible framework is to impose the existence of zero elements in the charged lepton and light neutrino mass matrices. Some of such zeroes may simply appear as a result of weak basis transformations~\cite{Branco:1999nb,Branco:2007nn} while others could be obtained from a symmetry principle~\cite{Grimus:2004hf,Felipe:2014vka}.

It is well established that neutrino mass matrices with more than two independent zero entries are not compatible with neutrino oscillation data, in the physical basis in which the charged leptons are diagonal and ordered ($e$,$\mu$,$\tau$). On the other hand, there are several Majorana neutrino patterns with two zeroes that are still viable~\cite{Frampton:2002yf,Fritzsch:2011qv,Meloni:2012sx,Meloni:2014yea,Cebola:2015dwa}. The latter contain eight real parameters of which five are physical (three unphysical phases can be removed). Another example of predictive textures with five physical parameters are the so-called hybrid textures~\cite{Kaneko:2005yz}, which have one texture zero and two equal nonzero elements in the physical basis. Most of these hybrid textures are also compatible with current data~\cite{Liu:2013oxa,Cebola:2015dwa}. Following the same line of reasoning, it is interesting to look for other neutrino textures which have the same predictability or which are even more predictive, i.e., contain less than five physical parameters. Since increasing the number of zeroes is not permitted, a straightforward possibility is to allow for correlations (equalities) of some elements in the effective neutrino mass matrix.

Our knowledge of neutrino masses and leptonic mixing has expanded in the last years thanks to solar, atmospheric, reactor and accelerator neutrino experiments~\cite{Capozzi:2013csa,Forero:2014bxa,Gonzalez-Garcia:2015qrr}, as well as to cosmological observations~\cite{Ade:2015xua}. Leptonic CP violation~\cite{Branco:2011zb} remains an important open question and it should be probed in near-future neutrino experiments~\cite{Adams:2013qkq}. For light Majorana neutrinos, there are nine specific observables that are relevant in the low-energy neutrino analysis: three neutrino masses, three mixing angles, one Dirac CP phase and two Majorana phases. These parameters should be determined from the physical parameters that characterise a given neutrino mass matrix.

An efficient and reliable approach to confront any neutrino pattern with the observational data is the $\chi^2$-analysis, which we shall use in this work. We shall look for predictive flavour patterns with equal elements (and possibly additional zeroes) in the effective Majorana neutrino mass matrix, and test their compatibility with data at $3\sigma\, (99.73\%)$ and $1\sigma\, (68.27\%)$ confidence levels (CL). 

The paper is organised as follows. In section~\ref{sec:textures}, we describe the minimal neutrino mass matrix textures and present their classification according to the number of physical parameters and correlated matrix elements. The number of solutions compatible with neutrino oscillation data at the $1\sigma$ and $3\sigma$ CL, for normal (NO) and inverted (IO) mass ordering, are also given. Section~\ref{sec:3param} is devoted to the search of maximally predictive textures. In particular, it is shown that there exist three unique patterns that can successfully explain neutrino observables at the $3\sigma$ CL with just three physical parameters. In section~\ref{sec:45param}, neutrino textures described by four and five physical parameters are discussed. The predictions for the lightest neutrino mass, effective mass parameter in neutrinoless double beta decays and CP-violating phases in the leptonic mixing are also given. Possible breaking paths of the viable three-parameter textures into four- and five-parameter textures compatible with neutrino oscillation data at $1\sigma$ are presented. Finally, our concluding remarks are given in section~\ref{sec:summary}.

\section{Minimal neutrino mass matrix textures}
\label{sec:textures}

We shall search for predictive flavour patterns of the effective Majorana neutrino mass matrix $\mathbf{m}_{\nu}$, based on the assumption that it contains equal matrix elements and, eventually, some vanishing elements. Our analysis is performed in the physical basis where the charged lepton mass matrix is diagonal, real and ordered, i.e., $\mathbf{m}_\ell=\mathrm{diag\,}(m_e,m_\mu,m_\tau)$. Moreover, we restrict our search to neutrino textures with a minimal number of physical parameters, $n<6$. By physical parameters we mean those that remain after removing all unphysical phases. 

In order to classify different textures according to the number of correlated matrix elements, we shall use the notation $N_{n_N}$ to specify that the matrix $\mathbf{m}_{\nu}$ contains $N$ equal elements and that such correlation appears $n_N$ times in the given texture. In this notation, $1_{n_1}$ simply denotes $n_1$ non-vanishing arbitrary elements, while the notation $0_{n_0}$ will be used to indicate that the neutrino mass matrix has $n_0$ independent texture zeroes. Since $\mathbf{m}_{\nu}$ is a $3\times3$ complex symmetric matrix, it has six independent entries. Clearly, any pattern with more than four equal elements is unrealistic. Thus, in general, we identify any given class of textures as $4_{n_4}3_{n_3}2_{n_2}1_{n_1}0_{n_0}$. The total number of textures for such a class can be determined through the combinatorial relation
\begin{equation}
\frac{6!}{n_4!\,(4!)^{n_4}\,n_3!\,(3!)^{n_3}\,n_2!\,(2!)^{n_2}\,n_1!\,n_0!}\,.
\end{equation}

\begin{table}[t]
	\centering
	\begin{tabular}{|ccccccc|}
		\hline
		$\;n\;$ & Texture class & No. of textures
		  & & \multicolumn{2}{c}{ Solutions $(1\sigma,3\sigma)$} &
		Table in the text\\
		    &        &      & & NO & IO & \\
		\hline
		2	& $4_10_2$ & $7^\ast$ $(\subset 15)$ & & - &  - & - \\
		\hline
		3	& $2_20_2$ & $21^\ast$ $(\subset 45)$ & & - &  - & -  \\
		& $3_11_10_2$ & $28^\ast$ $(\subset 60)$ & & - &  - & - \\
		& $3_12_10_1$ & 60 & & (0,1) & (0,2) & \ref{tab:T1D1Z1}\\
		& $3_2$ & 10 & & - &  - & - \\
		& $4_11_10_1$ & 30 & & - &  - & - \\
		& $4_12_1$ & 15 & & - &  - & - \\		 		
    	\hline
		4	& $2_11_20_2$ & $42^\ast$ $(\subset 90)$ & & (4,2) & (1,3) & \ref{tab:D1S2Z2}\\
		& $2_21_10_1$ & 90 & & (15,3) & (19,11) & \ref{tab:D2S1Z1-NO}, \ref{tab:D2S1Z1-IO}\\
		& $3_11_20_1$ & 60 & & (9,3) & (13,10) & \ref{tab:T1S2Z1}\\
		& $3_12_11_1$ & 60 & & (13,5) & (13,5) & \ref{tab:T1D1S1}\\
		& $4_11_2$ & 15 & & (5,0) & (3,1) & \ref{tab:Q1S2}\\
		\hline
		5	& $1_40_2$ & $7^\ast$ $(\subset 15)$ & & (6,0) & (3,2) & - \\
		& $2_11_30_1$ & 60 & & (31,4) & (36,7) & - \\
		& $2_3$ & 15 & & (4,0) & (1,1) & \ref{tab:D3}\\
		& $3_11_3$ & 20 & & (19,0) & (18,0) & \ref{tab:T1S3}\\
		\hline
	\end{tabular}
\caption{\label{tab:textures} Texture classification according to the number of physical parameters $n$. For each class, the number of different physical textures within the class is given. The number of solutions compatible with data at the $1\sigma$ and $3\sigma$ CL is also presented. Patterns marked with $\ast$ are particular cases of the well-known two-zero textures (denoted here as $1_4 0_2$) and, therefore, only the subset of potentially viable matrices are considered. The class $2_1 1_3 0_1$ corresponds to the so-called hybrid textures.}
\end{table}

In table~\ref{tab:textures} we present the texture classification according to the number of physical parameters $n$ and the correlated matrix elements. For any given texture class, the number of different physical textures within that class is also given. Patterns marked with $\ast$ are particular cases of the well-known two-zero textures~\cite{Frampton:2002yf} (denoted here as $1_4 0_2$). Only the subset of potentially viable matrices (i.e. patterns $\mathbf{A}_{1,2}$, $\mathbf{B}_{1,2,3,4}$ and $\mathbf{C}$ in the notation of~\cite{Frampton:2002yf}) has been considered~\cite{Fritzsch:2011qv,Meloni:2012sx,Meloni:2014yea,Cebola:2015dwa}. Note also that the class $2_1 1_3 0_1$ corresponds to the so-called hybrid textures with two equal elements and one zero~\cite{Kaneko:2005yz}.

In the next sections we confront all textures in table~\ref{tab:textures} with the low-energy observational neutrino data. We recall that in the framework of the standard model with three light neutrinos, leptonic mixing is described by the Pontecorvo-Maki-Nakagawa-Sakata unitary matrix $\mathbf{U}$, which relates the mass eigenstate neutrinos $\nu_i \ (i = 1,2,3)$ to the flavour eigenstate neutrinos $\nu_f \ (f=e,\mu,\tau)$~\cite{Pontecorvo:1957qd,Maki:1962mu,Pontecorvo:1967fh}. This matrix can be parametrised in terms of three mixing angles $\theta_{12}$, $\theta_{13}$, and $\theta_{23}$, a Dirac CP phase $\delta$ and two Majorana phases $\alpha_{21}$,  $\alpha_{31}$~\cite{Agashe:2014kda}. The neutrino mass matrix $\mathbf{m}_\nu$ is then diagonalised by the unitary transformation 
\begin{equation}
\mathbf{U}^{\dagger} \mathbf{m}_\nu \mathbf{U}^\ast=\mathrm{diag\,}(m_1,m_2,m_3)\,,
\end{equation}
where $m_i$ are the light neutrino physical masses. The mixing angles are easily reconstructed from $\mathbf{U}$ as
\begin{equation}
  \sin^2\theta_{12}=\dfrac{|U_{12}|^2}{1-|U_{13}|^2}\,,\quad
   \sin^2\theta_{23}=\dfrac{|U_{23}|^2}{1-|U_{13}|^2}\,, \quad
   \sin^2\theta _{13}=|U_{13}|^2\,.
\end{equation}
From the rephasing-invariant quartet $Q\equiv U_{12}U_{23}U^{\ast}_{13}U^{\ast}_{22}$, the Dirac phase $\delta$ can be found through the relations
\begin{subequations}
\begin{align}
\sin\delta&=\dfrac{8\,\text{Im}\,{(Q)}}{\cos\theta_{13}\sin2\theta_{12}\sin2\theta_{13}\sin2\theta_{23}}\,,\\
\cos\delta&=8\,\dfrac{\text{Re}\,{(Q)}+\cos^2\theta_{13}\sin^2\theta_{12}\sin^2\theta_{13}\sin^2\theta_{23}}{\cos\theta_{13}\sin2\theta_{12}\sin2\theta_{13}\sin2\theta_{23}}\,.
\end{align}
\end{subequations}
Finally, the Majorana phases are obtained as
\begin{equation}
\alpha_{21}=2\arg\left(U^{\ast}_{11}U_{12}\right)\,,\quad
\alpha_{31}=2\arg\left(U^{\ast}_{11}U_{13}\right)+2\delta\,.\\
\end{equation}

\begin{table}[t]
    \centering
	\begin{tabular}{|lcc|}
		\hline
		Parameter & Best fit $\pm$ $1\sigma$ & $3\sigma$\\
		\hline
		$\Delta m^2_{21}\: [10^{-5}\,\text{eV}^2]$
		& 7.60$^{+0.19}_{-0.18}$ & 7.11 -- 8.18\\[2mm] 
		$|\Delta m^2_{31}|\: [10^{-3}\,\text{eV}^2]$ (NO)
		& 2.48$^{+0.05}_{-0.07}$ & 2.30 -- 2.65\\
		\phantom{$|\Delta m^2_{31}|\: [10^{-3}\text{eV}^2]$ } (IO)
		& 2.38$^{+0.05}_{-0.06}$ & 2.20 -- 2.54\\[2mm] 
		$\sin^2\theta_{12} / 10^{-1}$
		& 3.23$\pm$0.16 & 2.78 -- 3.75\\[2mm] 
		$\sin^2\theta_{23} / 10^{-1}$ (NO)
		& 5.67$^{+0.32}_{-1.24}$ & 3.93 -- 6.43\\
		\phantom{$\sin^2\theta_{23} / 10^{-1}$ } (IO)
		& 5.73$^{+0.25}_{-0.39}$ & 4.03 -- 6.40\\[2mm]
		$\sin^2\theta_{13} / 10^{-2}$ (NO)
		& 2.26$\pm$0.12 & 1.90 -- 2.62\\
		\phantom{$\sin^2\theta_{13} / 10^{-2}$ } (IO)
		& 2.29$\pm$0.12 & 1.93 -- 2.65\\[2mm]
		$\delta/\pi$\quad (NO)
		& 1.41$^{+0.55}_{-0.40}$ & 0.0 -- 2.0\\
		\phantom{$\delta/\pi$ }\quad (IO)	
		& 1.48$\pm$0.31 & 0.0 -- 2.0\\	
		\hline
	\end{tabular}
		\caption{\label{tab:nudata} Neutrino oscillation parameters at the $68.27\%$ and $99.73\%$ CL~\cite{Forero:2014bxa}. The upper and lower rows in $\Delta m^2_{31}$, $\sin^2\theta_{23}$, $\sin^2\theta_{13}$,and $\delta$ correspond to NO and IO neutrino masses, respectively.}
\end{table}

Since the absolute scale of neutrino masses is unknown, there are two possible orderings of $m_i$, namely, a normal ordering (NO) with $m_1<m_2<m_3$ or an inverted ordering (IO) with $m_3<m_1<m_2$. The mass spectrum may vary from hierarchical to quasi-degenerate, yet cosmological observations place a stringent upper bound on the sum of the masses. Assuming three species of degenerate massive neutrinos and a $\Lambda$ cold dark matter model, the bound $\sum_i m_i < 0.23$~eV at 95\% CL has been obtained from a combined analysis of data~\cite{Ade:2015xua}. In what follows, we shall assume this bound in our numerical calculations. Notice also that in the physical basis ($\mathbf{m}_\ell$ diagonal) the absolute value of any matrix element of $\mathbf{m}_\nu$ is always smaller than the largest neutrino mass. Thus the above cosmological bound implies $|{m_{\nu}}_{ij}|\lesssim 80$~meV.

In our search for viable neutrino mass patterns at $3\sigma$ level, we shall consider five observables, namely, the mass-squared differences $\Delta m_{21}^2$ and $\Delta m_{31}^2$, and the mixing angles $\mathrm{sin}^2\theta_{12}$, $\mathrm{sin}^2\theta_{23}$, and $\mathrm{sin}^2\theta_{13}$. When testing the viability of the patterns at $1\sigma$ level, the Dirac CP phase $\delta$ is also included as an observable quantity. The current neutrino oscillation data are presented in table~\ref{tab:nudata} at the $1\sigma$ and $3\sigma$ CL~\cite{Forero:2014bxa}. If the predictions for the physical observables are within the ranges given in the table, a texture is considered to be consistent with observations. For such cases, we look for the corresponding predictions for the Dirac CP phase $\delta$, the lightest neutrino mass $m$, the effective neutrino mass parameter $m_{\beta\beta}=|\sum_i U_{ei}^2\,m_i|$, and the two CP-violating Majorana phases $\alpha_{21}, \alpha_{31}$. Note that, in the physical basis, one has $m_{\beta\beta}=|{m_\nu}_{ee}| \lesssim 80$~meV, for all textures considered in this work.

Concerning the atmospheric mixing angle $\theta_{23}$, it is worth noticing that its value is close to maximal ($\theta_{23}\approx\pi/4$). Starting from a given texture $\mathbf{m}_\nu$, one can always define a new texture $\mathbf{m}^{\prime}_\nu$, whose observables are directly related to those of $\mathbf{m}_\nu$,
\begin{equation}
\label{eq:p23trans}
\mathbf{m}^{\prime}_{\nu} = \mathbf{P}_{23}\,\mathbf{m}_{\nu}\,\mathbf{P}_{23}\,.
\end{equation}
Here $\mathbf{P}_{23}$ is the permutation matrix acting in the sector $(2,3)$. Said otherwise, $\mathbf{m}^{\prime}_{\nu}$ is obtained from $\mathbf{m}_{\nu}$ by interchanging its second and third rows and columns. The matrices $\mathbf{m}_\nu$ and $\mathbf{m}^{\prime}_{\nu}$ share the same mass spectrum, mixing angles $\theta_{12}$ and $\theta_{13}$, and Majorana phases $\alpha_{21}, \alpha_{31}$, while the mixing angle $\theta_{23}$ and the Dirac phase $\delta$ are transformed as
\begin{equation}
\label{eq:trans}
\theta_{23} \,\rightarrow\, \frac{\pi}{2}-\theta_{23}\,,\quad
\delta\,\rightarrow\, \delta-\pi\,.
\end{equation}
One can then conclude from  table~\ref{tab:nudata} that, if the given texture $\mathbf{m}_\nu$ is viable at $3\sigma$ CL, the transformed $\mathbf{m}^{\prime}_{\nu}$ would also be compatible with neutrino oscillation data, provided that $\sin^2\theta_{23}\leq 0.6$ (i.e. $\theta_{23}\leq 51^{\circ}$) in both NO and IO cases. 
 
We shall perform a $\chi^2$-analysis using the standard definition of the $\chi^2$-function in terms of the observable values and their deviations from the mean values. The minimisation of the $\chi^2$ function is carried out with respect to the neutrino observables using the MINUIT package~\cite{James:1975dr}, following the numerical strategy presented in~\cite{Cebola:2015dwa}. This procedure allows us to determine whether a given pattern is compatible at $1\sigma$ level. In order to obtain the allowed ranges at $3\sigma$ CL for the predictable quantities, we then randomly scan the full parameter space in the neutrino mass matrices. Since we do not impose any restriction on the complex matrix elements ${m_\nu}_{ij}$, our solutions exhibit an explicit symmetry on the CP phases: 
\begin{equation}
\label{eq:symmphase}
\delta\rightarrow-\delta\,,\quad \alpha_{21}\rightarrow-\alpha_{21}\,,\quad \alpha_{31}\rightarrow-\alpha_{31}\,.
\end{equation}
Since this symmetry does not change the value of any other observable or prediction, we use it to restrict the Dirac phase to the range $0\leq\delta\leq\pi$ when presenting our results. Obviously, values of $\delta$ in the range $\pi\leq\delta\leq2\pi$ are physical as well and may be even preferred by the experimental data~\cite{Abe:2013hdq} (cf. also table~\ref{tab:nudata}).

In the next sections, we seek for minimal neutrino textures that fulfil our criteria, i.e., that contain equal or vanishing elements and simultaneously have a minimal number of physical parameters. Since $\mathbf{m}_\nu$ is symmetric, we shall use the following sequential notation to label the position of the matrix elements of a given neutrino texture $\mathsf{T}$,
\begin{equation}
 \begin{pmatrix} 
 1 & 2 & 3 \\ 
 2 & 4 & 5 \\
 3 & 5 & 6
	\end{pmatrix}.
\end{equation}
We denote the position of any vanishing element labeled $i$ with a superscript, i.e. $\mathsf{T}^i$, while equal elements with labels $i,j,\dots, k$ are listed as subscripts, $\mathsf{T}_{ij\dots k}$. For instance, the notations $\mathsf{T}^{5,6}_{12,34}$ and $\mathsf{T}^{6}_{123,45}$ correspond to the textures
\begin{equation}
\mathsf{T}^{5,6}_{12,34} = 
\begin{pmatrix} 
a & a & b \\ 
a & b & 0 \\
b & 0 & 0
\end{pmatrix}\quad \text{and}
\quad
\mathsf{T}^{6}_{123,45} = 
\begin{pmatrix} 
a & a & a \\ 
a & b & b \\
a & b & 0
\end{pmatrix}\,,
\end{equation}
respectively.

\section{Textures with three parameters}
\label{sec:3param}

Following our $\chi^2$ numerical strategy, we have analysed all possible textures belonging to the classes listed in table~\ref{tab:textures}. The number of solutions compatible with neutrino oscillation data at the $1\sigma$ and $3\sigma$ CL are reported in this table.\footnote{Note that the number of solutions at $3\sigma$ refers to the number of textures that are consistent with data at $3\sigma$ but not at~$1\sigma$ CL.} From our search of the different pattern classes, no solution with only two physical parameters (class $4_10_2$) was found, neither for NO nor for IO neutrino spectrum. Thus, we conclude that under our assumptions (equal or vanishing elements) the minimal number of free physical parameters is 3. Moreover, from the table, one can see that the only viable class with $n=3$ is $3_1 2_1 0_1$. Out of the 60 textures within this class, only the patterns $\mathsf{T}^{2}_{136,45}$ for NO neutrino spectrum and $\mathsf{T}^{4}_{123,56}$ and $\mathsf{T}^{6}_{123,45}$ for IO spectrum are compatible with data at the $3\sigma$ level. None of the textures passed the $1\sigma$ CL test. 

For the texture $\mathsf{T}^{2}_{136,45}$ the effective neutrino mass matrix $\mathbf{m}_\nu$ has the following form
\begin{equation}
\mathbf{m}_\nu \equiv 
\begin{pmatrix} 
a & 0 & a \\ 
0 & b & b \\
a & b & a
\end{pmatrix}\,.
\end{equation}
The minimum of $\chi^2$ for this solution is $\chi^2_\text{min}=6.68$. An example that leads to this value is given by the choice of $a= 10.5731$~meV and $|b|=29$~meV, $\arg b=0.08284\pi$. This in turn yields the following low-energy neutrino parameters: 
\begin{equation}
\begin{aligned}
\Delta m^2_{21} &= 7.61\times10^{-5}\,\text{eV}^2,\quad |\Delta m^2_{31}| = 2.47\times10^{-3}\,\text{eV}^2, \\  
\sin^2\theta_{12} &= 3.14\times10^{-1}, \quad
\sin^2\theta_{23} = 6.31\times10^{-1}, \quad
\sin^2\theta_{13} = 2.44\times10^{-2}\,. 
\end{aligned}
\end{equation}
It also predicts 
\begin{equation}
\begin{aligned}
m_1 &= 12\,\text{meV},\quad m_2 = 15\,\text{meV},\quad m_3 = 51\,\text{meV},\quad m_{\beta\beta} = 11\,\text{meV},\\
\delta &= 0.41\pi,\quad \alpha_{21} = 1.5\pi,\quad \alpha_{31} = 0.70\pi\,.
\end{aligned}
\end{equation}

Similarly, for the texture $\mathsf{T}^{4}_{123,56}$ corresponding to
\begin{equation}
\mathbf{m}_\nu \equiv \begin{pmatrix} a & a & a \\ a & 0 & b \\
a & b & b
\end{pmatrix}\,,
\end{equation}
the minimum of $\chi^2$ is given by $\chi^2_\text{min}=5.96$. This minimum is obtained if we take, for instance, $a= 28.56$ meV and $|b|=18.8744$ meV, $\arg b=1.00845\pi$, which lead to the low-energy neutrino parameters
\begin{equation}
\begin{aligned}
\Delta m^2_{21} &= 7.59\times10^{-5}\,\text{eV}^2,\quad |\Delta m^2_{31}| = 2.38\times10^{-3}\,\text{eV}^2,  \\  
\sin^2\theta_{12} &= 3.25\times10^{-1}, \quad
\sin^2\theta_{23} = 5.74\times10^{-1}, \quad \sin^2\theta_{13} = 2.57\times10^{-2}\,,
\end{aligned}
\end{equation}
and the predictions
\begin{equation}
\begin{aligned}
m_1 &= 50\,\text{meV},\quad m_2 = 51\,\text{meV},\quad m_3 = 10\,\text{meV},\quad
m_{\beta\beta} = 29\,\text{meV},\\
\delta &= 0.32\pi,\quad
 \alpha_{21} = 1.3\pi,\quad \alpha_{31} = 0.49\pi.
\end{aligned}
\end{equation}

\begin{figure}[t]
	\centering 
	\includegraphics[width=\textwidth]{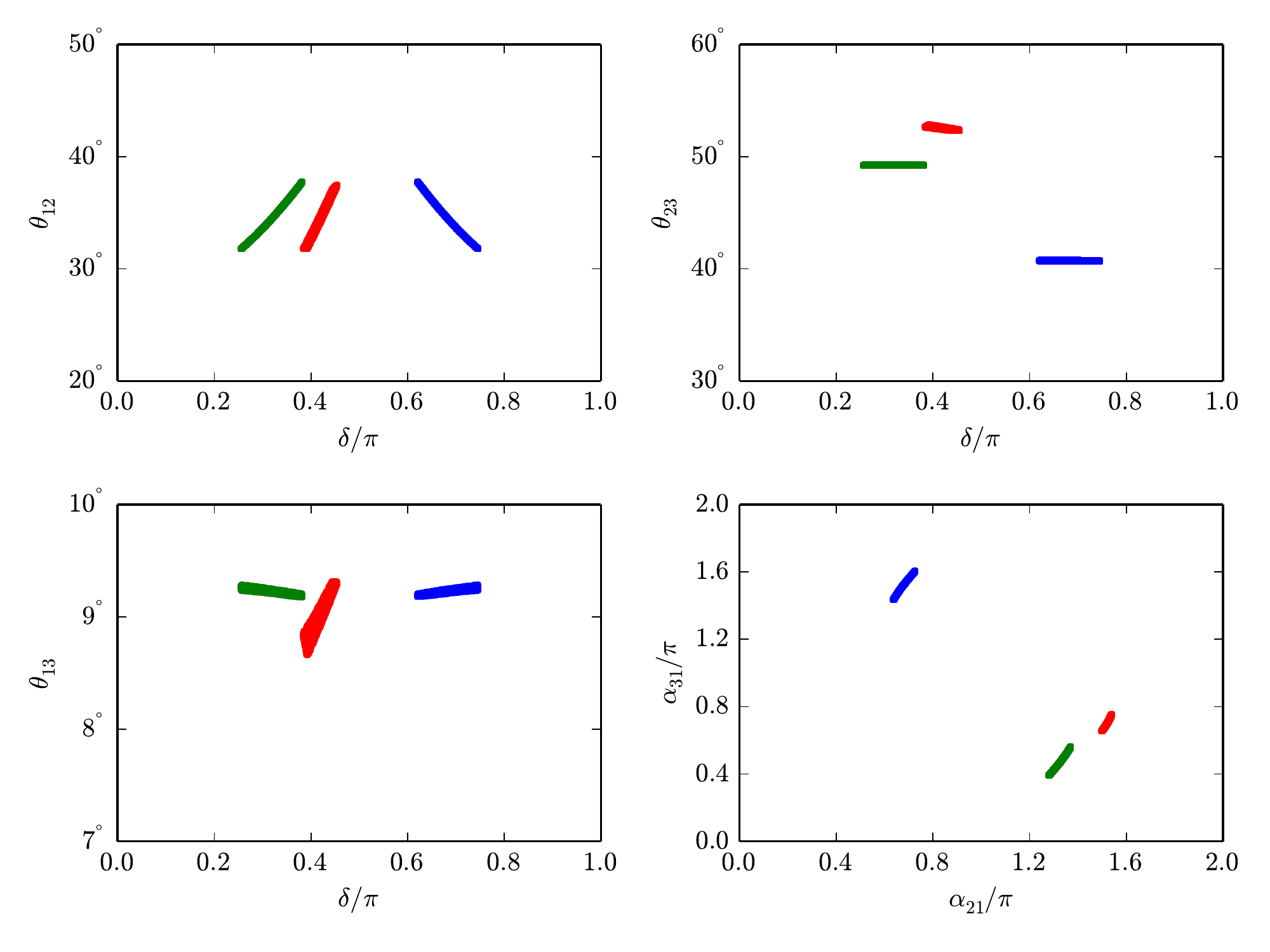}
	\caption{\label{fig1} Predictions for 3-parameter textures $\mathsf{T}^{2}_{136,45}$ (NO, red points), $\mathsf{T}^{4}_{123,56}$ (IO, green points) and $\mathsf{T}^{6}_{123,45}$ (IO, blue points).}
\end{figure}

Finally, for the texture $\mathsf{T}^{6}_{123,45}$ one has
\begin{equation}
\mathbf{m}_\nu \equiv 
\begin{pmatrix} 
a & a & a \\ 
a & b & b \\
a & b & 0
\end{pmatrix}\,.
\end{equation}
In this case, the minimum of $\chi^2$ corresponds to a somewhat larger value, $\chi^2_\text{min}=20.00$, which can be reached, e.g., with $a= 28.56$ meV and $|b|=18.8746$ meV, $\arg b=0.99155\pi$, leading to the following low-energy neutrino observables:
\begin{equation}
\begin{aligned}
\Delta m^2_{21} &= 7.59\times10^{-5}\,\text{eV}^2,\quad |\Delta m^2_{31}| = 2.38\times10^{-3}\,\text{eV}^2, \\  
\sin^2\theta_{12} &= 3.25\times10^{-1}, \quad
\sin^2\theta_{23} = 4.26\times10^{-1}, \quad
\sin^2\theta_{13} = 2.57\times10^{-2}\,. 
\end{aligned}
\end{equation}
The predicted mass spectrum and CP-violating phases are
\begin{equation}
\begin{aligned}
m_1 &= 50\,\text{meV},\quad m_2 = 51\,\text{meV},\quad m_3 = 10\,\text{meV},\quad m_{\beta\beta} = 29\,\text{meV},\\
\delta &= 0.68\pi,\quad \alpha_{21} = 0.67\pi,\quad \alpha_{31} = 1.5\pi.
\end{aligned}
\end{equation}

\begin{figure}[t]
	\begin{center}
	\subfigure[Class~$3_12_10_1$]{\label{fig2}\includegraphics[width=0.49\textwidth]{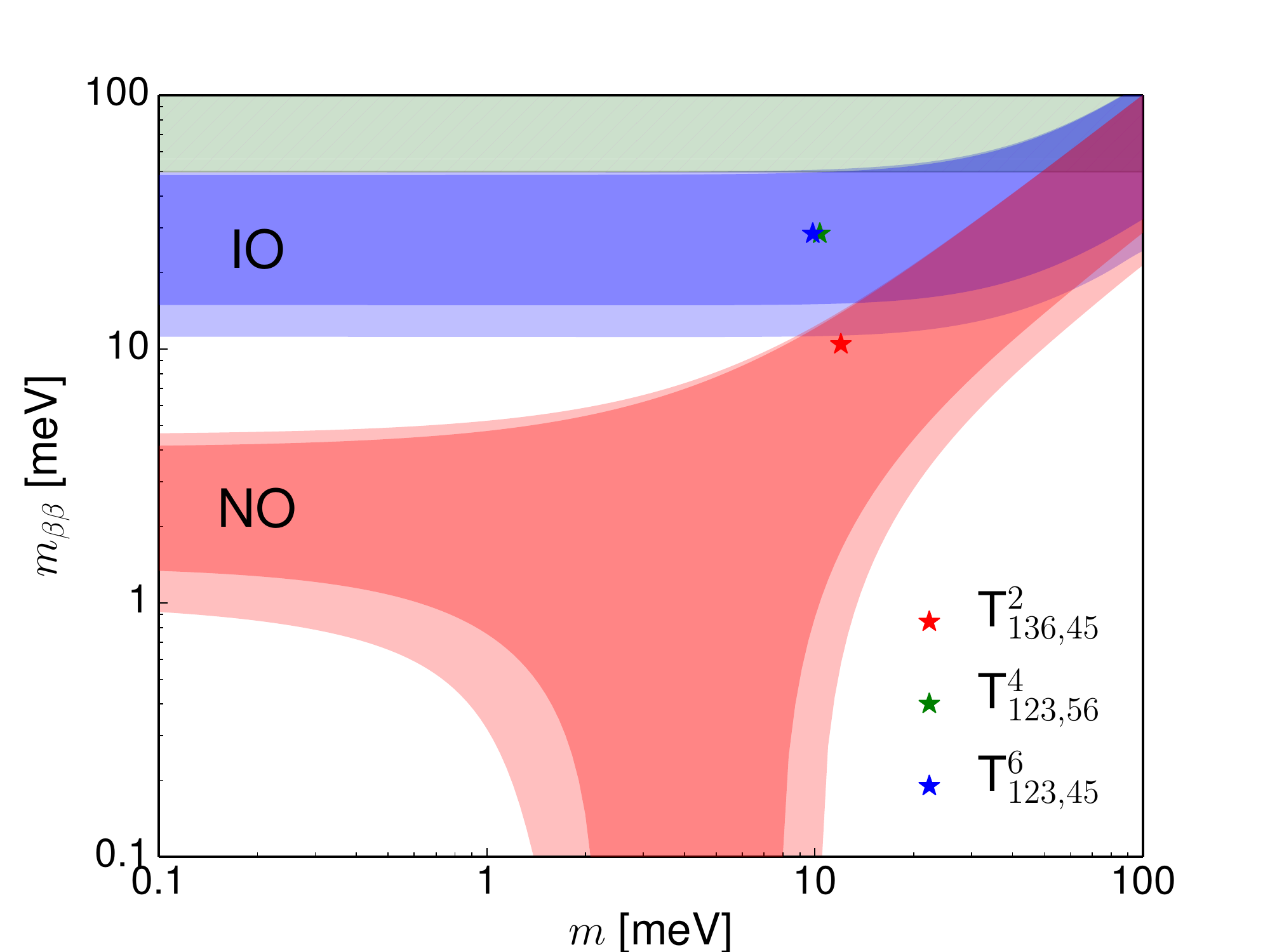}}
	 \subfigure[Class~$4_11_2$]{\label{fig3}\includegraphics[width=0.49\textwidth]{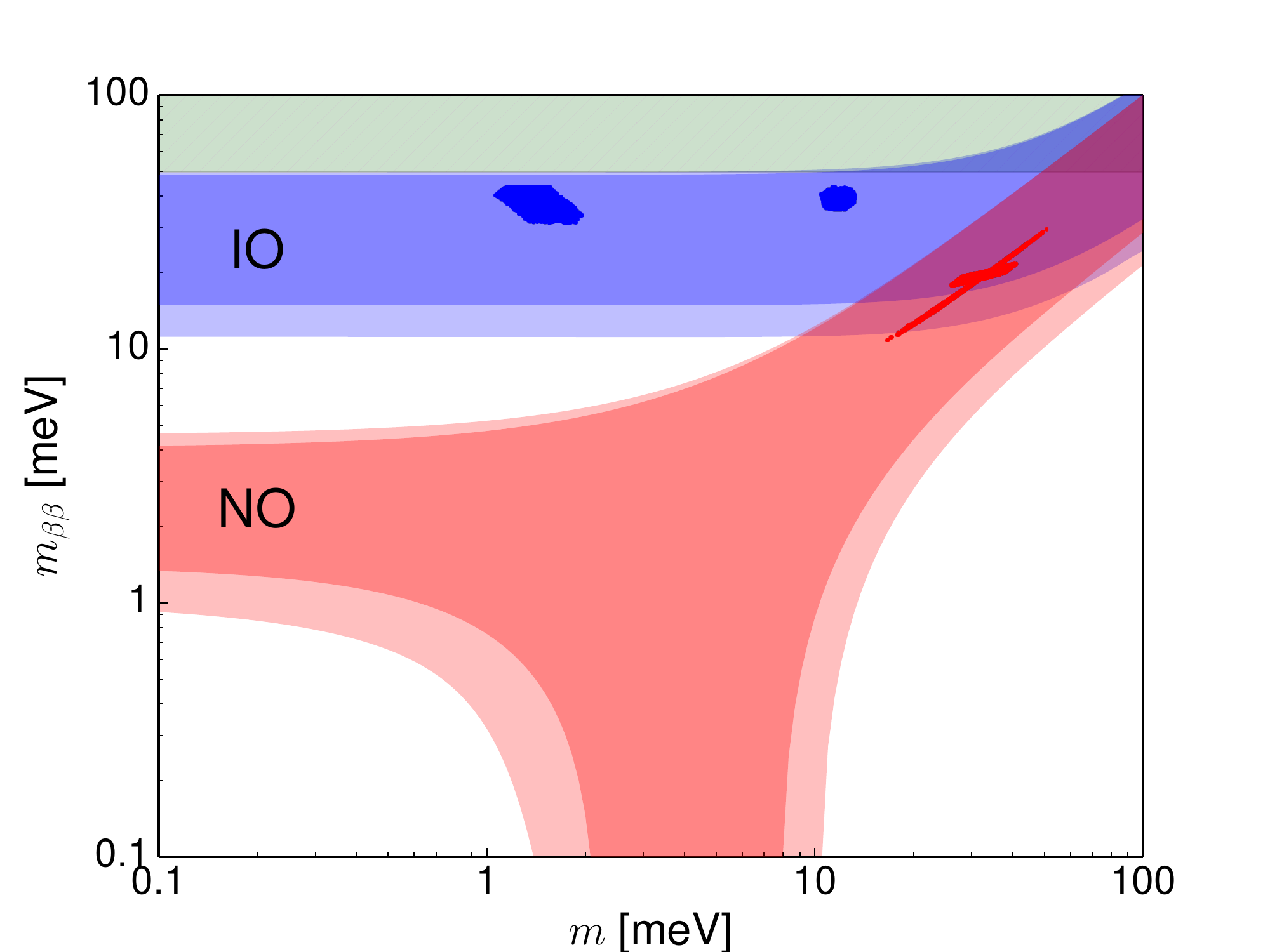}}
	
	\subfigure[Class~$2_11_20_2$]{\label{fig4}\includegraphics[width=0.49\textwidth]{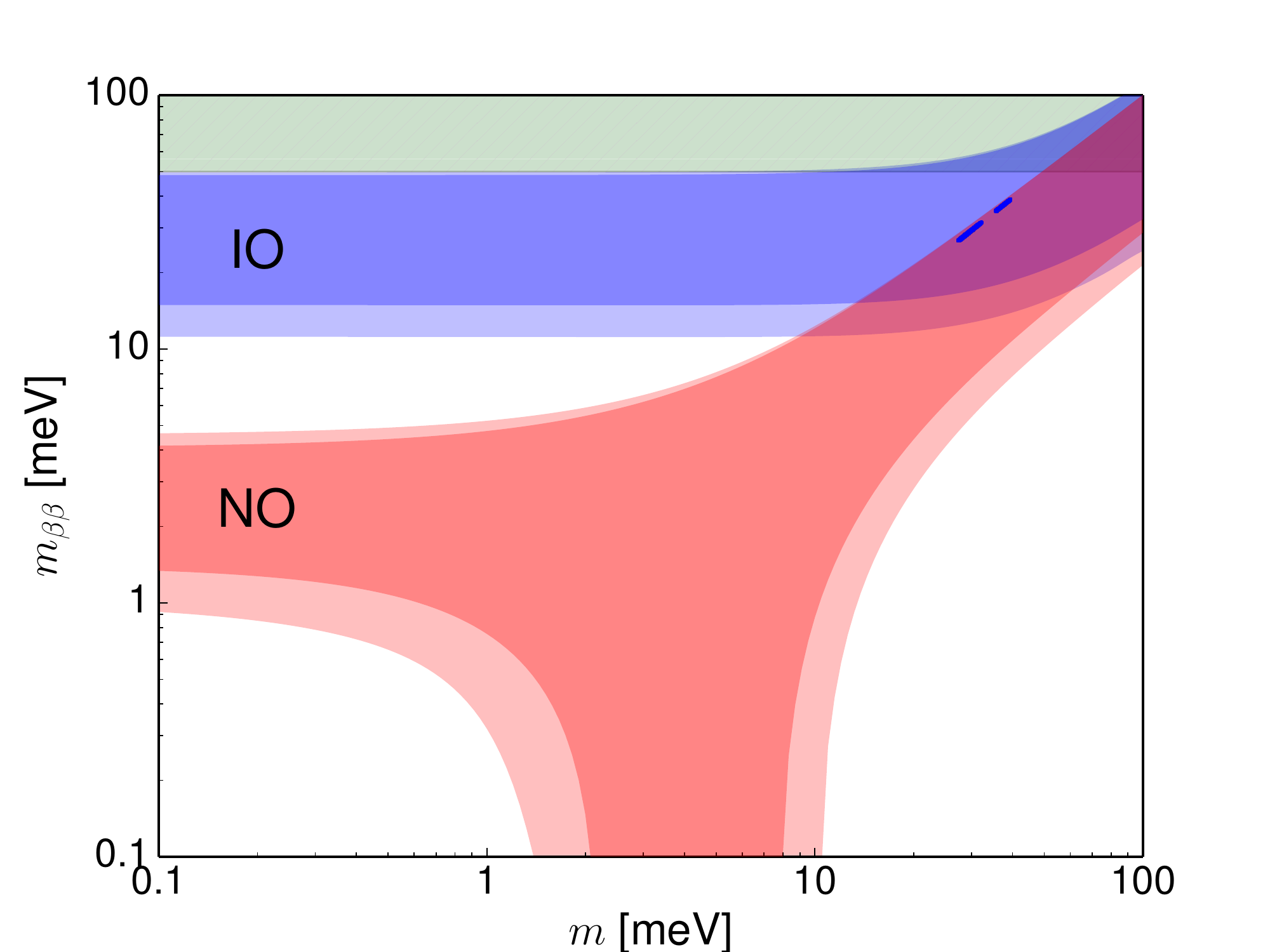}}
    \subfigure[Class~$2_3$]{\label{fig5}\includegraphics[width=0.49\textwidth]{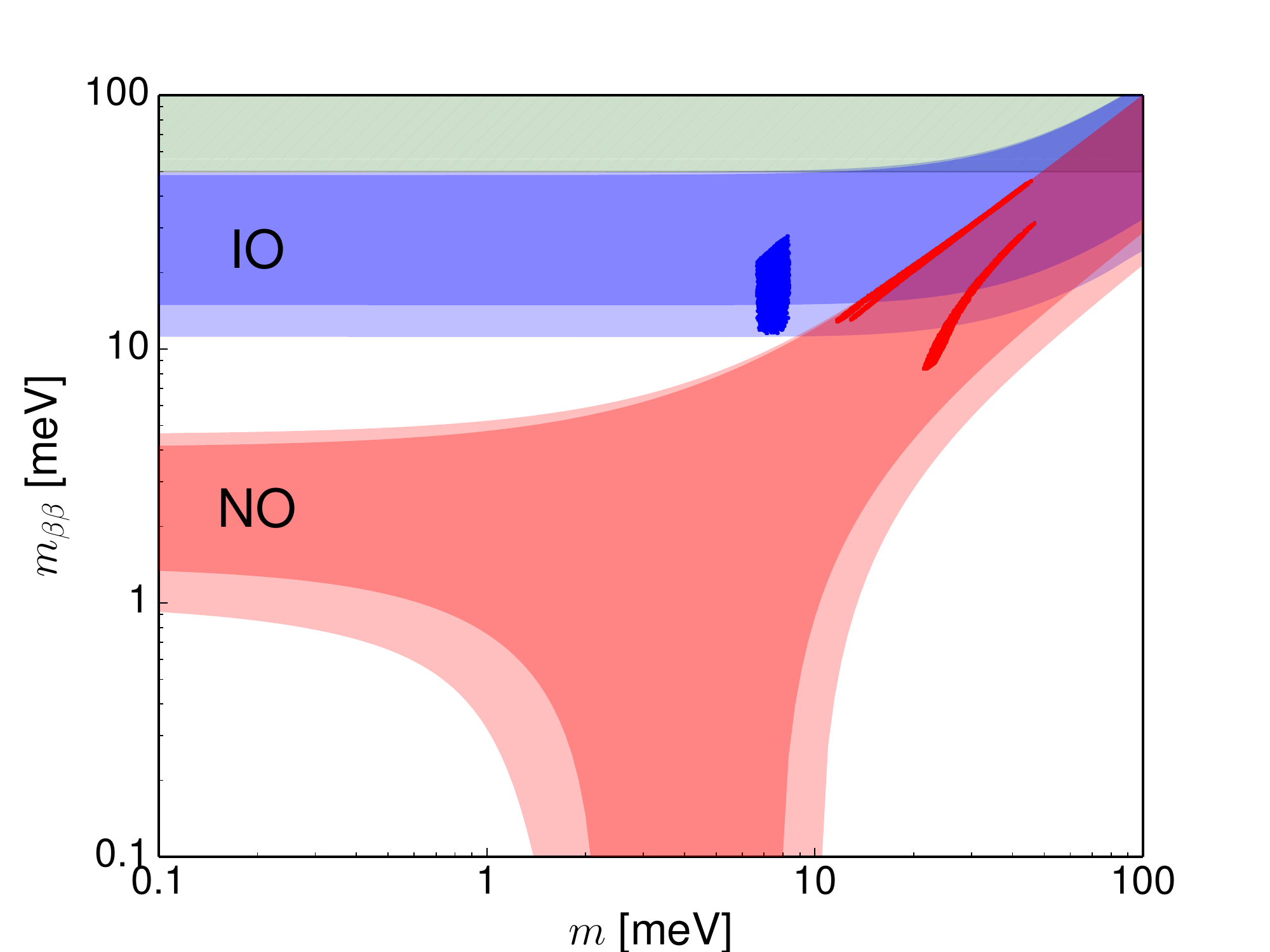}}
	\end{center}
	\caption{Predictions for $m_{\beta\beta}$ for different texture classes with NO (red colour) and IO (blue colour). The inner dark bands correspond to the $1\sigma$ neutrino oscillation parameters, while the lighter outer bands include the $3\sigma$ uncertainties. The green horizontal band corresponds to the expected sensitivity in near future of $0\nu\beta\beta$ experiments.}
\end{figure}

For each texture, we have also performed a random scan of the input parameters $a$ and $b$ in order to determine the viable ranges at $3\sigma$ for the quantities $\delta$, $m$, $m_{\beta\beta}$, $\alpha_{21}$ and $\alpha_{31}$. The results are collected in table~\ref{tab:T1D1Z1} of appendix~\ref{sec:tables} for the textures $\mathsf{T}^{2}_{136,45}$, $\mathsf{T}^{4}_{123,56}$ and $\mathsf{T}^{6}_{123,45}$. From the table one sees that neutrino observables are quite constrained in the three cases. In particular, for the texture $\mathsf{T}^{2}_{136,45}$ the Majorana phase $\alpha_{21}$ is predicted to be approximatively $3\pi/2$.

In figure~\ref{fig1}, we show the correlation of the Dirac phase $\delta$ with the three mixing angles, as well as the correlation between the Majorana phases. Notice that in virtue to the symmetry~\eqref{eq:symmphase} we restrict the values of $\delta$ to $[0,\pi]$. The effective mass parameter $m_{\beta\beta}$ as a function of the lightest neutrino mass $m$ is plotted in figure~\ref{fig2}. As can be seen in the figure, the predicted values for $m_{\beta\beta}$ are not far away from the values expected in the future CUORE~\cite{Alfonso:2015wka} and SuperNEMO~\cite{Remoto:2015wta} $0\nu\beta\beta$ experiments.

Notice that the textures $\mathsf{T}^{4}_{123,56}$ and $\mathsf{T}^{6}_{123,45}$ are related through the transformation given in equation~\eqref{eq:p23trans} and, therefore, their predictions are similar, except for the mixing angle $\theta_{23}$ and the Dirac phase $\delta$ that are transformed according to equation~\eqref{eq:trans}. We remark that for the Majorana phases $\alpha_{21}$ and $\alpha_{31}$ this conclusion may not be obvious in table~\ref{tab:T1D1Z1} and the numerical examples presented above, since after bringing the value of $\delta$ to the range $[0,\pi]$, the Majorana phases flip their signs. For the case of the texture $\mathsf{T}^{2}_{136,45}$, one has necessarily $\sin^2\theta_{23}>0.6$, i.e., $\theta_{23}>51^{\circ}$ (see figure~\ref{fig1}) and thus the transformed texture obtained from equation~\eqref{eq:p23trans} does not lead to a viable neutrino pattern.

\section{Textures with four and five parameters}
\label{sec:45param}

In this section, we consider the remaining classes in table~\ref{tab:textures} and look for solutions with four and five physical parameters compatible with low-energy neutrino data at $3\sigma$ CL, and eventually consistent with data at $1\sigma$. As it turns out, all classes (with $n=4,5$) contain viable solutions at $1\sigma$ CL. In appendix~\ref{sec:tables}, the predictions of the corresponding neutrino textures are summarised in tables~\ref{tab:D1S2Z2}--\ref{tab:Q1S2} for $n=4$ and tables~\ref{tab:D3} and \ref{tab:T1S3} for $n=5$.  As before, the values of $\delta$ are restricted to $[0,\pi]$. We remark that the five-parameter textures belonging to the classes $1_40_2$ and $2_11_30_1$ are not considered, since they have been studied in~\cite{Cebola:2015dwa}.

From the tables, we conclude that textures within the same class do not share necessarily the same predictions. Remarkably, some of the classes predict a quite constrained region in the plane $(m,m_{\beta\beta})$, namely the classes $4_11_2$, $2_11_20_2$ and $2_3$, as shown in figures~\ref{fig3}, \ref{fig4} and \ref{fig5}, respectively. It can also be noticed from the tables that there exist two disconnected $\delta$-regions for certain textures and, in some cases, the value of the Majorana phase $\alpha_{21}$ is essentially fixed. 

It is interesting to note that the class~$2_11_20_2$, which is more restrictive than the two-zero textures in~$1_40_2$, is also compatible with the data at both $1\sigma$ and $3\sigma$ CL. More precisely, three textures of the class~$1_40_2$, compatible with data at $1\sigma$, give rise to viable patterns in class~$2_11_20_2$. The textures $A_1$ and $A_2$ lead, for NO, to the textures $A_{1,45}$, $A_{1,46}$, $A_{2,46}$, $A_{2,56}$ (compatible at $1\sigma$ CL), and the textures $A_{1,56}$, $A_{2,45}$ (compatible only at $3\sigma$ CL). The remaining texture $C$ of class~$1_40_2$ leads, for IO, to four textures within class $2_11_20_2$: $C_{13}$ (compatible at $1\sigma$ CL) and $C_{12}$, $C_{25}$, $C_{35}$ (compatible only at $3\sigma$ CL). In labelling the above textures, we have used the notation of~\cite{Frampton:2002yf} for the $1_40_2$ textures and denoted the pair of equal elements in the corresponding textures of class $2_11_20_2$ with an extra subscript pair. Note also that the class~$2_11_20_2$ is the only viable class with textures having an additional correlation among the matrix elements of the two-zero textures of class~$1_40_2$. Particular cases of textures belonging to the class~$2_11_20_2$ have been previously considered in the literature. For example, the texture~$C_{25}$ was analysed in~\cite{Frigerio:2013uva}, while the textures~$A_{1,46}$ and $A_{2,46}$ with an additional constraint were found to be compatible with data at $3\sigma$ CL in~\cite{Grimus:2012zm}. We analysed all the textures of table~\ref{tab:D1S2Z2} in order to search for extra correlations among the matrix elements of $\mathbf{m}_{\nu}$. Besides the two textures given in~\cite{Grimus:2012zm}, we have found eight new textures that are consistent with observations at $3\sigma$~CL. Remarkably, there  are also four patterns with only three physical parameters that can accommodate the neutrino data at $1\sigma$~CL.  A summary of these results is presented in appendix~\ref{sec:extracorr}.

In section~\ref{sec:3param}, we have shown that the three-parameter textures $\mathsf{T}^{2}_{136,45}$ (for NO neutrino spectrum), $\mathsf{T}^{4}_{123,56}$ and $\mathsf{T}^{6}_{123,45}$ (for IO spectrum) are compatible with data at the $3\sigma$~CL. Assuming that these patterns could be implemented through symmetries, one may wonder whether they can be made compatible with neutrino oscillation data at the $1\sigma$ CL, once the symmetries are (softly) broken. Since the above patterns belong to the texture class $3_1 2_1 0_1$, four natural breaking paths can be envisaged: the breaking to the classes $3_1 2_1 1_1$, $3_1 1_2 0_1$, $2_2 1_1 0_1$ or $2_1 1_3 0_1$, obtained from the initial class $3_1 2_1 0_1$ by breaking the zero entry, the doublet or the triplet (one or two elements), respectively. Let us analyse each possibility in detail. 

Looking at tables~\ref{tab:T1S2Z1} and~\ref{tab:T1D1S1} we can easily verify that the breaking paths $\mathsf{T}^{2}_{136,45} \rightarrow \mathsf{T}_{136,45}$ and $\mathsf{T}^{2}_{136,45} \rightarrow \mathsf{T}^2_{136}$ are not viable either for NO or IO mass spectrum. From table~\ref{tab:D2S1Z1-NO} we also conclude that the pattern $\mathsf{T}^{2}_{36,45}$ is not compatible with data at $1\sigma$. On the other hand, the textures $\mathsf{T}^{2}_{13,45}$ and $\mathsf{T}^{2}_{16,45}$ turn out to be viable for NO spectrum. Finally, the breaking $\mathsf{T}^{2}_{136,45} \rightarrow \mathsf{T}^{2}_{45}$ results in a hybrid texture that is compatible with data for both NO and IO neutrino mass spectra~\cite{Cebola:2015dwa}. We note that in the notation of~\cite{Cebola:2015dwa} the texture $\mathsf{T}^{2}_{45}$ corresponds to the matrix $\widehat{\mathbf{D}}_{1(12)}$.

Let us now consider the texture $\mathsf{T}^{4}_{123,56}$. In this case, breaking it to the classes $3_1 2_1 1_1$ and $3_1 1_2 0_1$ yields the viable patterns $\mathsf{T}_{123,56}$ and $\mathsf{T}^{4}_{123}$ for IO spectrum, as can be seen from tables~\ref{tab:T1S2Z1} and~\ref{tab:T1D1S1}. Similarly, the breaking path to the class $2_2 1_1 0_1$, which contains two pairs of equal elements, leads to three textures compatible with data, $\mathsf{T}^{4}_{12,56}$, $\mathsf{T}^{4}_{13,56}$, and $\mathsf{T}^{4}_{23,56}$, all of them for an IO neutrino spectrum (cf. table~\ref{tab:D2S1Z1-IO}). Furthermore, there is also a viable pattern within the hybrid class $2_1 1_3 0_1$, namely, the matrix $\mathsf{T}^{4}_{56}$ (or $\widehat{\mathbf{D}}_{2(22)}$ in the notation of~\cite{Cebola:2015dwa}).

Finally, for the texture $\mathsf{T}^{6}_{123,45}$, only the paths $\mathsf{T}^{6}_{123,45} \rightarrow \mathsf{T}_{123,45}$ and $\mathsf{T}^{6}_{123,45} \rightarrow \mathsf{T}^{6}_{45}$ (or $\widehat{\mathbf{D}}_{1(33)}$ in the notation of~\cite{Cebola:2015dwa}) turn out to be viable, both of them for an IO neutrino mass spectrum. A summary of the breaking paths for the three-parameter textures is presented in table~\ref{tab:breakingpaths}.

\begin{table}
\centering 
\begin{tabular}{|c|cccc|}
\hline
Texture & $3_12_11_1$ & $3_11_20_1$ & $2_21_10_1$ & $2_11_30_1$ \\
\hline
$\mathsf{T}^{2}_{136,45}$ & $\mathsf{T}_{136,45}$ {\scriptsize( - )} & $\mathsf{T}^{2}_{136}$ {\scriptsize( - )} & $\mathsf{T}^{2}_{13,45}$ {\scriptsize(NO)} & $\mathsf{T}^{2}_{45}$ {\scriptsize(NO, IO)}\\[0.2mm]
& & & $\mathsf{T}^{2}_{16,45}$ {\scriptsize(NO)} & \\[0.2mm]
& & & $\mathsf{T}^{2}_{36,45}$ {\scriptsize( - )} & \\[0.2mm]
$\mathsf{T}^{4}_{123,56}$ & $\mathsf{T}_{123,56}$ {\scriptsize(IO)} & $\mathsf{T}^{4}_{123}$ {\scriptsize(IO)} & $\mathsf{T}^{4}_{12,56}$ {\scriptsize(IO)} & $\mathsf{T}^{4}_{56}$ {\scriptsize(IO)} \\[0.2mm]
& & & $\mathsf{T}^{4}_{13,56}$ {\scriptsize(IO)} & \\[0.2mm]
& & & $\mathsf{T}^{4}_{23,56}$ {\scriptsize(IO)} & \\[0.2mm]
$\mathsf{T}^{6}_{123,45}$ & $\mathsf{T}_{123,45}$ {\scriptsize(IO)} & $\mathsf{T}^{6}_{123}$ {\scriptsize( - )} & $\mathsf{T}^{6}_{12,45}$ {\scriptsize( - )} & $\mathsf{T}^{6}_{45}$ {\scriptsize(IO)} \\[0.2mm]
& & & $\mathsf{T}^{6}_{13,45}$ {\scriptsize( - )} & \\[0.2mm]
& & & $\mathsf{T}^{6}_{23,45}$ {\scriptsize( - )} & \\[0.2mm]
\hline
\end{tabular}
\caption{\label{tab:breakingpaths} Neutrino mass textures arising from the breaking of either the texture zero or the equality of elements in the three-parameter matrices of table~\ref{tab:T1D1Z1}. Their compatibility with data at $1\sigma$ is also shown.}
\end{table}

\section{Conclusions}
\label{sec:summary}

Despite the overwhelming success of the standard model, the nature of the flavour structure of the quark and lepton sectors is still unknown. One of the common approaches to address this puzzle is to assume certain constraints on the mass matrices in order to reduce the number of free parameters.

In this paper, we looked for predictive flavour patterns of the effective Majorana neutrino mass matrix that are compatible with current neutrino oscillation data at the $3\sigma$ CL. Our study was based on the assumption that the effective neutrino mass matrix contains some restrictions in the physical basis where the charged leptons are diagonal and properly ordered. In particular, we assumed the existence of equal matrix elements or the  presence of texture zeroes (see table~\ref{tab:textures}). Following a standard $\chi^2$-analysis, we found three unique patterns that are described by just three physical parameters and can successfully explain current neutrino data at the $3\sigma$ CL. The predictions of these solutions for the neutrino observables, namely, the lightest neutrino mass, the effective mass parameter in neutrinoless double beta decays, and the CP-violating phases, are summarised in table~\ref{tab:T1D1Z1}. 

We have also studied neutrino textures described by four and five physical parameters. Several viable patterns were found to be compatible with the data at $1\sigma$ and $3\sigma$ CL. Their predictions are given in tables~\ref{tab:D1S2Z2}--\ref{tab:Q1S2} for $n=4$ and tables~\ref{tab:D3} and \ref{tab:T1S3} for $n=5$. From our study we can conclude that current neutrino data do not allow to discriminate, among the viable predictive neutrino mass matrix textures, the patterns that more appropriately describe the observations. Upcoming neutrino experiments and more precise future measurements could shed some light on this issue.

The neutrino textures considered in this work were taken as simple and economical anz\"{a}tze. It would be interesting to see if such predictive textures could arise from a symmetry principle. For instance, it has been shown that the two-zero textures in class $1_40_2$ can be obtained from continuous or discrete flavour symmetries (see e.g.~\cite{Fritzsch:2011qv,Cebola:2013hta,Felipe:2014vka}), while the particular texture~$\mathsf{T}^{4,6}_{25}$ of class $2_11_20_2$ can be constructed through a flavour symmetry based on the order-twelve quaternion group $Q_6$~\cite{Frigerio:2013uva}. Similarly, the hybrid textures belonging to class $2_11_30_1$ can be realised in a type-II seesaw framework by implementing certain discrete symmetries~\cite{Liu:2013oxa}. Another aspect that deserves further study is the stability of the textures under the renormalisation group evolution. 

\acknowledgments
The authors acknowledge support from Funda\c c\~ ao para a Ci\^ encia e a Tecnologia (FCT, Portugal) through the projects CERN/FIS-NUC/0010/2015 and UID/FIS/00777/2013. L.C. is supported by FCT under the contract PD/BD/113479/2015.

\appendix

\section{Summary tables for classes with four and five parameters}
\label{sec:tables}

In this appendix we summarise the predictions for the Dirac CP phase $\delta$, the lightest neutrino mass $m$, the effective neutrino mass parameter $m_{\beta\beta}$, and the two CP-violating Majorana phases $\alpha_{21}, \alpha_{31}$ for all the viable textures listed in table~\ref{tab:textures}. The results for $n=3$, $n=4$, and $n=5$ physical parameters are given in table~\ref{tab:T1D1Z1}, tables~\ref{tab:D1S2Z2}--\ref{tab:Q1S2}, and tables~\ref{tab:D3}, \ref{tab:T1S3}, respectively. 
\bigskip\bigskip 


\begin{table}[h]
	\centering 
	\scalebox{0.9}{
\begin{tabular}{|cccccccc|}
		\hline
		Spectrum &Texture & $\delta/\pi$ & $m$ [meV] & $m_{\beta\beta}$ [meV] & $\alpha_{21}/\pi$ & $\alpha_{31}/\pi$ & $1\sigma$ CL\\
		\hline
		NO & $\mathsf{T}^{2}_{136,45}$ & $0.38$ -- $0.45$ & $11$ -- $13$ & $9.9$ -- $11$ & $\approx 1.5$ & $0.66$ -- $0.75$ & - 
		\\[0.2mm]
		\hline
		IO & $\mathsf{T}^{4}_{123,56}$ & $0.26$ -- $0.38$ & $9.7$ -- $11$ & $27$ -- $30$ & $1.3$ -- $1.4$ & $0.39$ -- $0.56$ & - 
		\\[0.2mm]
		& $\mathsf{T}^{6}_{123,45}$ & $0.62$ -- $0.74$ & $9.7$ -- $10$ & $27$ -- $30$ & $0.63$ -- $0.72$ & $1.4$ -- $1.6$ & - 
		\\[0.2mm]
		\hline
	\end{tabular}}
	\caption{\label{tab:T1D1Z1} Neutrino mass textures in class $3_12_10_1$ compatible with data at $3\sigma$ CL.}
\end{table}


\begin{table}
\centering
\scalebox{0.9}{
\begin{tabular}{|cccccccc|}
\hline
Spectrum & Texture & $\delta/\pi$ & $m$ [meV] & $m_{\beta\beta}$ [meV] & $\alpha_{21}/\pi$ & $\alpha_{31}/\pi$ & $1\sigma$ CL\\
\hline
NO & $\mathsf{T}^{1,3}_{45}$ & $0.0$ -- $1.0$ & $2.6$ -- $8.9$ & 0 & $1.0$ -- $1.1$ & $1.0$ -- $2.0$ & -\\[0.2mm]
 & $\mathsf{T}^{1,2}_{45}$ & $0.26$ -- $0.62$ & $4.1$ -- $5.9$ & 0 & $0.88$ -- $0.93$ & $0.22$ -- $0.54$ & $\checkmark$\\[0.2mm]
 & $\mathsf{T}^{1,3}_{46}$ & $0.0$ -- $1.0$ & $3.9$ -- $6.7$ & 0 & $1.0$ -- $1.1$ & $1.0$ -- $2.0$ & $\checkmark$\\[0.2mm]
 & $\mathsf{T}^{1,2}_{46}$ & $0.0$ -- $1.0$ & $3.9$ -- $6.6$ & 0 & $0.89$ -- $1.0$ & $0.0$ -- $1.0$ & $\checkmark$\\[0.2mm]
 & $\mathsf{T}^{1,3}_{56}$ & $0.31$ -- $0.75$ & $4.1$ -- $5.9$ & 0 & $\approx 1.1$ & $1.4$ -- $1.8$ & $\checkmark$\\[0.2mm]
 & $\mathsf{T}^{1,2}_{56}$ & $0.0$ -- $1.0$ & $2.7$ -- $8.9$ & 0 & $0.89$ -- $1.0$ & $0.0$ -- $1.0$ & -\\[0.2mm]
\hline
IO & $\mathsf{T}^{4,6}_{12}$ & $0.27$ -- $0.38$ & $36$ -- $39$ & $36$ -- $39$ & $1.3$ -- $1.4$ & $0.43$ -- $0.58$ & -\\
 & & $0.63$ -- $0.80$ & $29$ -- $32$ & $29$ -- $32$ & $0.68$ -- $0.80$ & $1.5$ -- $1.7$ & \\[0.2mm]
 & $\mathsf{T}^{4,6}_{13}$ & $0.61$ -- $0.73$ & $36$ -- $39$ & $36$ -- $39$ & $0.61$ -- $0.69$ & $1.4$ -- $1.6$ & $\checkmark$\\[0.2mm]
 & $\mathsf{T}^{4,6}_{25}$ & $0.27$ -- $0.38$ & $35$ -- $38$ & $35$ -- $38$ & $1.3$ -- $1.4$ & $0.41$ -- $0.57$ & -\\
 & & $0.65$ -- $0.84$ & $27$ -- $31$ & $27$ -- $30$ & $0.69$ -- $0.84$ & $1.5$ -- $1.8$ & \\[0.2mm]
 & $\mathsf{T}^{4,6}_{35}$ & $0.62$ -- $0.73$ & $35$ -- $38$ & $35$ -- $38$ & $0.62$ -- $0.71$ & $1.4$ -- $1.6$ & -\\[0.2mm]
\hline
\end{tabular}}
\caption{\label{tab:D1S2Z2} Neutrino mass textures in class $2_11_20_2$ compatible with data at $3\sigma$ CL.}
\end{table}

\begin{table}[h]
\centering
\scalebox{0.9}{
\begin{tabular}{|cccccccc|}
\hline
Spectrum & Texture & $\delta/\pi$ & $m$ [meV] & $m_{\beta\beta}$ [meV] & $\alpha_{21}/\pi$ & $\alpha_{31}/\pi$ & $1\sigma$ CL\\
\hline
NO & $\mathsf{T}^{3}_{12,45}$ & $0.063$ -- $1.0$ & $6.2$ -- $15$ & $7.4$ -- $13$ & $0.33$ -- $1.6$ & $0.10$ -- $1.7$ & $\checkmark$\\[0.2mm]
& $\mathsf{T}^{3}_{12,46}$ & $0.0$ -- $1.0$ & $7.3$ -- $19$ & $8.0$ -- $15$ & $0.38$ -- $1.6$ & $0.0$ -- $2.0$ & $\checkmark$\\[0.2mm]
& $\mathsf{T}^{3}_{12,56}$ & $0.072$ -- $0.47$ & $7.2$ -- $11$ & $8.1$ -- $10$ & $0.35$ -- $0.48$ & $0.61$ -- $1.2$ & $\checkmark$\\
&  & $0.79$ -- $1.0$ & $11$ -- $14$ & $9.6$ -- $12$ & $0.41$ -- $1.7$ & $0.12$ -- $1.7$ & \\[0.2mm]
& $\mathsf{T}^{2}_{13,45}$ & $0.0$ -- $0.82$ & $8.4$ -- $13$ & $8.5$ -- $12$ & $0.39$ -- $1.6$ & $0.25$ -- $1.8$ & $\checkmark$\\[0.2mm]
& $\mathsf{T}^{2}_{13,46}$ & $0.0$ -- $1.0$ & $7.3$ -- $19$ & $8.0$ -- $15$ & $0.36$ -- $1.6$ & $0.0$ -- $2.0$ & $\checkmark$\\[0.2mm]
& $\mathsf{T}^{2}_{13,56}$ & $0.0$ -- $0.20$ & $9.0$ -- $17$ & $8.8$ -- $14$ & $0.36$ -- $1.6$ & $0.31$ -- $2.0$ & $\checkmark$\\
&  & $0.55$ -- $0.96$ & $6.4$ -- $15$ & $7.5$ -- $13$ & $1.5$ -- $1.6$ & $0.98$ -- $1.3$ & \\[0.2mm]
& $\mathsf{T}^{3}_{14,26}$ & $0.48$ -- $0.61$ & $22$ -- $26$ & $23$ -- $27$ & $0.081$ -- $0.12$ & $0.98$ -- $1.2$ & -\\[0.2mm]
& $\mathsf{T}^{2}_{14,36}$ & $0.36$ -- $0.56$ & $22$ -- $29$ & $22$ -- $29$ & $1.8$ -- $1.9$ & $0.75$ -- $1.0$ & $\checkmark$\\[0.2mm]
& $\mathsf{T}^{3}_{14,56}$ & $0.19$ -- $0.33$ & $16$ -- $44$ & $16$ -- $45$ & $0.051$ -- $0.31$ & $0.52$ -- $0.89$ & $\checkmark$\\
&  & $0.83$ -- $0.89$ & $17$ -- $44$ & $17$ -- $45$ & $0.056$ -- $0.32$ & $1.5$ -- $1.6$ & \\[0.2mm]
& $\mathsf{T}^{2}_{14,56}$ & $0.12$ -- $0.17$ & $22$ -- $47$ & $22$ -- $47$ & $1.8$ -- $1.9$ & $0.41$ -- $0.49$ & $\checkmark$\\
&  & $0.75$ -- $0.81$ & $21$ -- $49$ & $22$ -- $49$ & $\approx 1.9$ & $1.3$ -- $1.5$ & \\[0.2mm]
& $\mathsf{T}^{3}_{15,46}$ & $0.0$ -- $1.0$ & $19$ -- $70$ & $16$ -- $70$ & $0.024$ -- $1.8$ & $0.0$ -- $2.0$ & $\checkmark$\\[0.2mm]
& $\mathsf{T}^{2}_{15,46}$ & $0.0$ -- $1.0$ & $19$ -- $71$ & $16$ -- $71$ & $0.16$ -- $2.0$ & $0.0$ -- $2.0$ & $\checkmark$\\[0.2mm]
& $\mathsf{T}^{3}_{16,24}$ & $0.47$ -- $0.62$ & $21$ -- $25$ & $22$ -- $26$ & $0.14$ -- $0.19$ & $1.0$ -- $1.2$ & -\\[0.2mm]
& $\mathsf{T}^{3}_{16,45}$ & $0.19$ -- $0.34$ & $14$ -- $42$ & $15$ -- $42$ & $0.076$ -- $0.19$ & $0.53$ -- $0.82$ & $\checkmark$\\
&  & $0.82$ -- $0.88$ & $15$ -- $45$ & $16$ -- $45$ & $0.088$ -- $0.21$ & $1.4$ -- $1.6$ & \\[0.2mm]
& $\mathsf{T}^{2}_{16,45}$ & $0.11$ -- $0.81$ & $11$ -- $41$ & $9.7$ -- $42$ & $1.5$ -- $1.9$ & $0.41$ -- $1.5$ & $\checkmark$\\[0.2mm]
& $\mathsf{T}^{3}_{25,46}$ & $0.076$ -- $0.87$ & $22$ -- $70$ & $18$ -- $68$ & $1.7$ -- $1.9$ & $0.0$ -- $2.0$ & $\checkmark$\\[0.2mm]
& $\mathsf{T}^{2}_{35,46}$ & $0.13$ -- $0.92$ & $22$ -- $71$ & $18$ -- $70$ & $0.095$ -- $0.34$ & $0.0$ -- $2.0$ & $\checkmark$\\[0.2mm]
& $\mathsf{T}^{2}_{36,45}$ & $0.22$ -- $0.53$ & $11$ -- $15$ & $9.2$ -- $14$ & $1.4$ -- $1.6$ & $0.45$ -- $0.88$ & -\\[0.2mm]
\hline
\end{tabular}}
\caption{\label{tab:D2S1Z1-NO} Neutrino mass textures in class $2_21_10_1$ compatible with data at $3\sigma$ CL for NO spectrum.}
\end{table}

\begin{table}
\centering
\scalebox{0.9}{
\begin{tabular}{|cccccccc|}
\hline
Spectrum & Texture & $\delta/\pi$ & $m$ [meV] & $m_{\beta\beta}$ [meV] & $\alpha_{21}/\pi$ & $\alpha_{31}/\pi$ & $1\sigma$ CL\\
\hline
IO & $\mathsf{T}^{5}_{12,34}$ & $0.0$ -- $1.0$ & $30$ -- $39$ & $29$ -- $39$ & $0.63$ -- $1.4$ & $0.43$ -- $1.6$ & $\checkmark$\\[0.2mm]
& $\mathsf{T}^{5}_{12,36}$ & $0.0$ -- $0.34$ & $30$ -- $36$ & $31$ -- $36$ & $0.63$ -- $1.3$ & $0.43$ -- $1.6$ & $\checkmark$\\
&  & $0.57$ -- $0.99$ & $30$ -- $36$ & $31$ -- $36$ & $1.2$ -- $1.4$ & $1.3$ -- $1.6$ & \\[0.2mm]
& $\mathsf{T}^{6}_{12,45}$ & $0.44$ -- $0.91$ & $9.7$ -- $14$ & $27$ -- $30$ & $0.62$ -- $0.72$ & $1.3$ -- $1.7$ & -\\[0.2mm]
& $\mathsf{T}^{5}_{12,46}$ & $0.12$ -- $0.29$ & $29$ -- $38$ & $28$ -- $38$ & $0.62$ -- $0.81$ & $0.44$ -- $0.67$ & $\checkmark$\\
&  & $0.67$ -- $0.89$ & $29$ -- $37$ & $28$ -- $37$ & $1.2$ -- $1.4$ & $1.2$ -- $1.6$ & \\[0.2mm]
& $\mathsf{T}^{4}_{12,56}$ & $0.0$ -- $0.71$ & $9.8$ -- $17$ & $26$ -- $30$ & $0.67$ -- $1.4$ & $0.16$ -- $1.5$ & $\checkmark$\\[0.2mm]
& $\mathsf{T}^{5}_{13,24}$ & $0.0$ -- $0.35$ & $32$ -- $36$ & $32$ -- $36$ & $0.63$ -- $1.4$ & $0.42$ -- $1.6$ & $\checkmark$\\
&  & $0.60$ -- $0.97$ & $32$ -- $36$ & $32$ -- $37$ & $1.3$ -- $1.4$ & $1.4$ -- $1.6$ & \\[0.2mm]
& $\mathsf{T}^{5}_{13,26}$ & $0.0$ -- $1.0$ & $30$ -- $38$ & $29$ -- $38$ & $0.61$ -- $1.4$ & $0.41$ -- $1.6$ & $\checkmark$\\[0.2mm]
& $\mathsf{T}^{6}_{13,45}$ & $0.42$ -- $0.94$ & $9.8$ -- $14$ & $27$ -- $30$ & $0.63$ -- $0.74$ & $1.3$ -- $1.8$ & -\\[0.2mm]
& $\mathsf{T}^{5}_{13,46}$ & $0.095$ -- $0.27$ & $31$ -- $40$ & $30$ -- $40$ & $0.60$ -- $0.75$ & $0.38$ -- $0.66$ & $\checkmark$\\
&  & $0.73$ -- $0.88$ & $30$ -- $41$ & $30$ -- $40$ & $1.2$ -- $1.4$ & $1.4$ -- $1.6$ & \\[0.2mm]
& $\mathsf{T}^{4}_{13,56}$ & $0.0$ -- $0.68$ & $9.7$ -- $17$ & $27$ -- $32$ & $0.66$ -- $1.4$ & $0.24$ -- $1.4$ & $\checkmark$\\[0.2mm]
& $\mathsf{T}^{6}_{14,23}$ & $0.22$ -- $1.0$ & $0.53$ -- $27$ & $13$ -- $17$ & $0.83$ -- $1.1$ & $0.0$ -- $2.0$ & -\\[0.2mm]
& $\mathsf{T}^{5}_{14,23}$ & $0.042$ -- $0.90$ & $12$ -- $65$ & $12$ -- $65$ & $0.45$ -- $1.6$ & $0.29$ -- $1.7$ & $\checkmark$\\[0.2mm]
& $\mathsf{T}^{6}_{14,25}$ & $0.0$ -- $0.90$ & $49$ -- $65$ & $27$ -- $32$ & $0.77$ -- $1.2$ & $0.0049$ -- $2.0$ & $\checkmark$\\[0.2mm]
& $\mathsf{T}^{5}_{14,26}$ & $0.077$ -- $0.24$ & $27$ -- $38$ & $27$ -- $37$ & $0.63$ -- $0.83$ & $0.45$ -- $0.69$ & $\checkmark$\\
&  & $0.60$ -- $0.84$ & $27$ -- $37$ & $27$ -- $36$ & $1.2$ -- $1.4$ & $1.2$ -- $1.6$ & \\[0.2mm]
& $\mathsf{T}^{5}_{14,36}$ & $0.068$ -- $0.32$ & $26$ -- $37$ & $24$ -- $37$ & $0.63$ -- $0.90$ & $0.40$ -- $0.88$ & $\checkmark$\\
&  & $0.48$ -- $0.83$ & $26$ -- $37$ & $24$ -- $37$ & $1.1$ -- $1.4$ & $1.1$ -- $1.5$ & \\[0.2mm]
& $\mathsf{T}^{6}_{15,23}$ & $0.0$ -- $0.87$ & $14$ -- $66$ & $12$ -- $66$ & $0.44$ -- $1.6$ & $0.0$ -- $2.0$ & -\\[0.2mm]
& $\mathsf{T}^{4}_{15,23}$ & $0.13$ -- $1.0$ & $15$ -- $65$ & $12$ -- $65$ & $0.41$ -- $1.6$ & $0.0$ -- $2.0$ & $\checkmark$\\[0.2mm]
& $\mathsf{T}^{5}_{16,23}$ & $0.097$ -- $0.95$ & $13$ -- $66$ & $12$ -- $65$ & $0.44$ -- $1.6$ & $0.29$ -- $1.7$ & -\\[0.2mm]
& $\mathsf{T}^{4}_{16,23}$ & $0.0$ -- $0.83$ & $0.53$ -- $27$ & $13$ -- $17$ & $0.85$ -- $1.2$ & $0.0$ -- $2.0$ & $\checkmark$\\[0.2mm]
& $\mathsf{T}^{5}_{16,24}$ & $0.18$ -- $0.93$ & $24$ -- $36$ & $21$ -- $36$ & $0.64$ -- $1.4$ & $0.47$ -- $1.6$ & $\checkmark$\\[0.2mm]
& $\mathsf{T}^{5}_{16,34}$ & $0.15$ -- $0.35$ & $29$ -- $40$ & $29$ -- $39$ & $0.62$ -- $0.78$ & $0.39$ -- $0.69$ & $\checkmark$\\
&  & $0.77$ -- $0.93$ & $29$ -- $40$ & $29$ -- $39$ & $1.2$ -- $1.4$ & $1.3$ -- $1.5$ & \\[0.2mm]
& $\mathsf{T}^{4}_{16,35}$ & $0.25$ -- $0.99$ & $54$ -- $65$ & $29$ -- $32$ & $0.79$ -- $1.2$ & $0.012$ -- $2.0$ & -\\[0.2mm]
& $\mathsf{T}^{6}_{23,45}$ & $0.62$ -- $0.90$ & $7.7$ -- $11$ & $22$ -- $30$ & $0.62$ -- $0.91$ & $1.4$ -- $1.9$ & -\\[0.2mm]
& $\mathsf{T}^{4}_{23,56}$ & $0.095$ -- $0.39$ & $7.7$ -- $11$ & $22$ -- $30$ & $1.1$ -- $1.4$ & $0.14$ -- $0.58$ & $\checkmark$\\[0.2mm]
& $\mathsf{T}^{5}_{24,36}$ & $0.0$ -- $0.33$ & $31$ -- $35$ & $32$ -- $35$ & $0.65$ -- $0.75$ & $0.42$ -- $0.65$ & $\checkmark$\\
&  & $0.62$ -- $0.96$ & $32$ -- $35$ & $32$ -- $35$ & $1.2$ -- $1.4$ & $1.3$ -- $1.6$ & \\[0.2mm]
& $\mathsf{T}^{4}_{25,36}$ & $0.35$ -- $0.42$ & $12$ -- $14$ & $36$ -- $38$ & $\approx 1.5$ & $0.58$ -- $0.68$ & -\\[0.2mm]
& $\mathsf{T}^{3}_{25,46}$ & $0.20$ -- $0.77$ & $19$ -- $64$ & $47$ -- $79$ & $1.8$ -- $1.9$ & $0.0$ -- $2.0$ & -\\[0.2mm]
& $\mathsf{T}^{5}_{26,34}$ & $0.0$ -- $1.0$ & $32$ -- $38$ & $32$ -- $35$ & $0.64$ -- $1.4$ & $0.44$ -- $1.6$ & -\\[0.2mm]
& $\mathsf{T}^{4}_{26,35}$ & $0.10$ -- $0.65$ & $13$ -- $19$ & $36$ -- $39$ & $\approx 1.5$ & $0.47$ -- $0.81$ & -\\[0.2mm]
& $\mathsf{T}^{2}_{35,46}$ & $0.23$ -- $0.80$ & $19$ -- $66$ & $47$ -- $80$ & $0.094$ -- $0.18$ & $0.0$ -- $2.0$ & $\checkmark$\\[0.2mm]
\hline
\end{tabular}}
\caption{\label{tab:D2S1Z1-IO} Neutrino mass textures in class $2_21_10_1$ compatible with data at $3\sigma$ CL for IO spectrum.}
\end{table}	

\begin{table}[h]
\centering
\scalebox{0.9}{
\begin{tabular}{|cccccccc|}
\hline
Spectrum & Texture & $\delta/\pi$ & $m$ [meV] & $m_{\beta\beta}$ [meV] & $\alpha_{21}/\pi$ & $\alpha_{31}/\pi$ & $1\sigma$ CL\\
\hline
NO & $\mathsf{T}^{3}_{125}$ & $0.19$ -- $0.63$ & $20$ -- $24$ & $16$ -- $18$ & $\approx 1.6$ & $1.5$ -- $1.9$ & -\\[0.2mm]
 & $\mathsf{T}^{6}_{134}$ & $0.0$ -- $0.90$ & $53$ -- $60$ & $34$ -- $39$ & $0.54$ -- $1.5$ & $0.45$ -- $1.6$ & -\\[0.2mm]
 & $\mathsf{T}^{2}_{135}$ & $0.32$ -- $0.85$ & $20$ -- $24$ & $15$ -- $18$ & $0.38$ -- $0.46$ & $0.096$ -- $0.47$ & $\checkmark$\\[0.2mm]
 & $\mathsf{T}^{2}_{136}$ & $0.38$ -- $0.46$ & $11$ -- $13$ & $9.7$ -- $11$ & $\approx 1.5$ & $0.66$ -- $0.76$ & -\\[0.2mm]
 & $\mathsf{T}^{3}_{145}$ & $0.16$ -- $0.25$ & $29$ -- $39$ & $29$ -- $40$ & $0.052$ -- $0.17$ & $0.55$ -- $0.62$ & $\checkmark$\\
 & & $0.80$ -- $0.93$ & $29$ -- $39$ & $29$ -- $40$ & $0.060$ -- $0.18$ & $1.5$ -- $1.7$ & \\[0.2mm]
 & $\mathsf{T}^{2}_{145}$ & $0.087$ -- $0.21$ & $30$ -- $41$ & $30$ -- $41$ & $\approx 1.9$ & $0.39$ -- $0.53$ & $\checkmark$\\
 & & $0.72$ -- $0.83$ & $30$ -- $40$ & $30$ -- $41$ & $\approx 1.9$ & $1.3$ -- $1.5$ & \\[0.2mm]
 & $\mathsf{T}^{3}_{146}$ & $0.096$ -- $0.94$ & $16$ -- $69$ & $17$ -- $69$ & $0.045$ -- $0.23$ & $0.33$ -- $1.7$ & $\checkmark$\\[0.2mm]
 & $\mathsf{T}^{2}_{146}$ & $0.062$ -- $0.91$ & $16$ -- $69$ & $16$ -- $69$ & $1.8$ -- $2.0$ & $0.25$ -- $1.7$ & $\checkmark$\\[0.2mm]
 & $\mathsf{T}^{3}_{156}$ & $0.16$ -- $0.24$ & $28$ -- $39$ & $29$ -- $39$ & $0.081$ -- $0.11$ & $0.52$ -- $0.63$ & $\checkmark$\\
 & & $0.81$ -- $0.93$ & $29$ -- $38$ & $29$ -- $39$ & $0.10$ -- $0.12$ & $1.5$ -- $1.6$ & \\[0.2mm]
 & $\mathsf{T}^{2}_{156}$ & $0.060$ -- $0.18$ & $29$ -- $38$ & $28$ -- $39$ & $1.8$ -- $1.9$ & $0.31$ -- $0.50$ & $\checkmark$\\
 & & $0.76$ -- $0.85$ & $28$ -- $38$ & $28$ -- $39$ & $1.8$ -- $1.9$ & $\approx 1.4$ & \\[0.2mm]
 & $\mathsf{T}^{6}_{234}$ & $0.0$ -- $1.0$ & $54$ -- $71$ & $48$ -- $70$ & $0.12$ -- $2.0$ & $0.86$ -- $1.3$ & $\checkmark$\\[0.2mm]
 & $\mathsf{T}^{4}_{236}$ & $0.0$ -- $1.0$ & $54$ -- $72$ & $48$ -- $70$ & $0.036$ -- $1.9$ & $0.74$ -- $1.1$ & $\checkmark$\\[0.2mm]
\hline
IO & $\mathsf{T}^{6}_{123}$ & $0.0$ -- $1.0$ & $9.4$ -- $57$ & $27$ -- $43$ & $0.62$ -- $1.4$ & $0.34$ -- $1.8$ & -\\[0.2mm]
 & $\mathsf{T}^{5}_{123}$ & $0.0$ -- $0.40$ & $31$ -- $36$ & $33$ -- $36$ & $0.64$ -- $1.4$ & $0.43$ -- $1.6$ & $\checkmark$\\
 & & $0.51$ -- $1.0$ & $32$ -- $36$ & $33$ -- $36$ & $0.64$ -- $1.4$ & $0.43$ -- $1.6$ & \\[0.2mm]
 & $\mathsf{T}^{4}_{123}$ & $0.0$ -- $1.0$ & $9.5$ -- $56$ & $27$ -- $43$ & $0.62$ -- $1.4$ & $0.24$ -- $1.6$ & $\checkmark$\\[0.2mm]
 & $\mathsf{T}^{6}_{125}$ & $0.0$ -- $1.0$ & $25$ -- $37$ & $28$ -- $38$ & $0.61$ -- $1.4$ & $0.37$ -- $1.8$ & $\checkmark$\\[0.2mm]
 & $\mathsf{T}^{4}_{125}$ & $0.0$ -- $1.0$ & $25$ -- $39$ & $27$ -- $38$ & $0.60$ -- $1.4$ & $0.25$ -- $1.8$ & $\checkmark$\\[0.2mm]
 & $\mathsf{T}^{5}_{126}$ & $0.17$ -- $0.41$ & $29$ -- $37$ & $27$ -- $37$ & $0.62$ -- $0.85$ & $0.45$ -- $0.72$ & $\checkmark$\\
 & & $0.72$ -- $0.93$ & $29$ -- $37$ & $27$ -- $37$ & $1.1$ -- $1.4$ & $1.2$ -- $1.6$ & \\[0.2mm]
 & $\mathsf{T}^{5}_{134}$ & $0.061$ -- $0.23$ & $30$ -- $39$ & $29$ -- $39$ & $0.61$ -- $0.80$ & $0.38$ -- $0.74$ & $\checkmark$\\
 & & $0.62$ -- $0.84$ & $30$ -- $39$ & $29$ -- $39$ & $1.2$ -- $1.4$ & $1.3$ -- $1.6$ & \\[0.2mm]
 & $\mathsf{T}^{6}_{135}$ & $0.0$ -- $1.0$ & $30$ -- $39$ & $31$ -- $38$ & $0.60$ -- $1.4$ & $0.40$ -- $1.7$ & $\checkmark$\\[0.2mm]
 & $\mathsf{T}^{4}_{135}$ & $0.0$ -- $1.0$ & $26$ -- $37$ & $28$ -- $38$ & $0.60$ -- $1.4$ & $0.23$ -- $1.6$ & $\checkmark$\\[0.2mm]
 & $\mathsf{T}^{6}_{145}$ & $0.36$ -- $0.96$ & $13$ -- $20$ & $12$ -- $17$ & $0.86$ -- $1.1$ & $0.0011$ -- $2.0$ & -\\[0.2mm]
 & $\mathsf{T}^{4}_{156}$ & $0.055$ -- $0.66$ & $13$ -- $20$ & $12$ -- $17$ & $0.86$ -- $1.1$ & $0.0$ -- $2.0$ & -\\[0.2mm]
 & $\mathsf{T}^{6}_{234}$ & $0.0$ -- $0.32$ & $30$ -- $66$ & $48$ -- $78$ & $0.21$ -- $1.9$ & $0.81$ -- $1.3$ & -\\
 & & $0.60$ -- $1.0$ & $22$ -- $66$ & $43$ -- $78$ & $0.22$ -- $1.9$ & $0.85$ -- $1.3$ & \\[0.2mm]
 & $\mathsf{T}^{5}_{234}$ & $0.0$ -- $0.45$ & $29$ -- $38$ & $29$ -- $39$ & $0.63$ -- $1.4$ & $0.42$ -- $1.6$ & $\checkmark$\\
 & & $0.58$ -- $1.0$ & $29$ -- $38$ & $29$ -- $39$ & $0.68$ -- $1.4$ & $0.46$ -- $1.6$ & \\[0.2mm]
 & $\mathsf{T}^{6}_{235}$ & $0.0$ -- $1.0$ & $15$ -- $49$ & $33$ -- $38$ & $0.51$ -- $1.4$ & $0.23$ -- $1.8$ & -\\[0.2mm]
 & $\mathsf{T}^{4}_{235}$ & $0.0$ -- $1.0$ & $14$ -- $49$ & $33$ -- $38$ & $0.59$ -- $1.5$ & $0.24$ -- $1.8$ & $\checkmark$\\[0.2mm]
 & $\mathsf{T}^{5}_{236}$ & $0.0$ -- $0.97$ & $25$ -- $37$ & $27$ -- $37$ & $0.62$ -- $1.4$ & $0.42$ -- $1.5$ & $\checkmark$\\[0.2mm]
 & $\mathsf{T}^{4}_{236}$ & $0.0$ -- $1.0$ & $15$ -- $66$ & $38$ -- $78$ & $0.12$ -- $1.8$ & $0.54$ -- $1.2$ & -\\[0.2mm]
 & $\mathsf{T}^{5}_{246}$ & $0.13$ -- $0.31$ & $27$ -- $38$ & $27$ -- $37$ & $0.62$ -- $0.84$ & $0.45$ -- $0.71$ & $\checkmark$\\
 & & $0.63$ -- $0.89$ & $27$ -- $38$ & $27$ -- $37$ & $1.2$ -- $1.4$ & $1.2$ -- $1.6$ & \\[0.2mm]
 & $\mathsf{T}^{4}_{256}$ & $0.17$ -- $0.59$ & $13$ -- $16$ & $35$ -- $38$ & $1.4$ -- $1.5$ & $0.46$ -- $0.79$ & -\\[0.2mm]
 & $\mathsf{T}^{5}_{346}$ & $0.098$ -- $0.31$ & $29$ -- $40$ & $29$ -- $39$ & $0.61$ -- $0.80$ & $0.39$ -- $0.73$ & $\checkmark$\\
 &  & $0.72$ -- $0.87$ & $29$ -- $39$ & $28$ -- $39$ & $1.2$ -- $1.4$ & $1.3$ -- $1.6$ & \\[0.2mm]
 & $\mathsf{T}^{4}_{356}$ & $0.21$ -- $0.58$ & $13$ -- $16$ & $36$ -- $39$ & $\approx 1.5$ & $0.50$ -- $0.79$ & -\\[0.2mm]
 & $\mathsf{T}^{3}_{456}$ & $0.48$ -- $0.51$ & $0.96$ -- $1.4$ & $45$ -- $49$ & $\approx 1.9$ & $1.9$ -- $2.0$ & -\\[0.2mm]
 & $\mathsf{T}^{2}_{456}$ & $0.49$ -- $0.52$ & $0.96$ -- $1.4$ & $45$ -- $49$ & $0.096$ -- $0.12$ & $0.024$ -- $0.065$ & -\\[0.2mm]
\hline
\end{tabular}}
\caption{\label{tab:T1S2Z1} Neutrino mass textures in class $3_11_20_1$ compatible with data at $3\sigma$ CL.}
\end{table}

\begin{table}[h]
\centering
\scalebox{0.9}{
\begin{tabular}{|cccccccc|}
\hline
Spectrum & Texture & $\delta/\pi$ & $m$ [meV] & $m_{\beta\beta}$ [meV] & $\alpha_{21}/\pi$ & $\alpha_{31}/\pi$ & $1\sigma$ CL\\
\hline
NO & $\mathsf{T}_{124,36}$ & $0.0$ -- $1.0$ & $31$ -- $35$ & $17$ -- $22$ & $0.58$ -- $1.4$ & $0.25$ -- $1.7$ & $\checkmark$\\[0.2mm]
 & $\mathsf{T}_{125,46}$ & $0.0$ -- $0.98$ & $21$ -- $42$ & $16$ -- $21$ & $0.63$ -- $1.6$ & $0.36$ -- $1.8$ & $\checkmark$\\[0.2mm]
 & $\mathsf{T}_{126,34}$ & $0.0$ -- $1.0$ & $29$ -- $35$ & $16$ -- $24$ & $0.55$ -- $1.4$ & $0.12$ -- $1.9$ & $\checkmark$\\[0.2mm]
 & $\mathsf{T}_{126,45}$ & $0.0$ -- $1.0$ & $20$ -- $31$ & $10$ -- $13$ & $0.60$ -- $1.4$ & $0.0$ -- $2.0$ & $\checkmark$\\[0.2mm]
 & $\mathsf{T}_{134,26}$ & $0.0$ -- $1.0$ & $29$ -- $35$ & $16$ -- $24$ & $0.56$ -- $1.4$ & $0.14$ -- $1.9$ & $\checkmark$\\[0.2mm]
 & $\mathsf{T}_{134,56}$ & $0.29$ -- $1.0$ & $26$ -- $31$ & $11$ -- $13$ & $0.73$ -- $1.3$ & $0.0$ -- $2.0$ & -\\[0.2mm]
 & $\mathsf{T}_{135,46}$ & $0.0$ -- $1.0$ & $21$ -- $44$ & $16$ -- $22$ & $0.38$ -- $1.4$ & $0.20$ -- $1.6$ & $\checkmark$\\[0.2mm]
 & $\mathsf{T}_{136,24}$ & $0.0$ -- $1.0$ & $31$ -- $35$ & $17$ -- $22$ & $0.59$ -- $1.4$ & $0.27$ -- $1.7$ & $\checkmark$\\[0.2mm]
 & $\mathsf{T}_{136,45}$ & $0.25$ -- $0.58$ & $11$ -- $14$ & $9.5$ -- $12$ & $\approx 1.5$ & $0.57$ -- $0.80$ & -\\[0.2mm]
 & $\mathsf{T}_{145,23}$ & $0.29$ -- $0.32$ & $29$ -- $39$ & $29$ -- $40$ & $1.9$ -- $2.0$ & $0.41$ -- $0.56$ & $\checkmark$\\
 & & $0.61$ -- $0.63$ & $29$ -- $39$ & $30$ -- $40$ & $1.9$ -- $2.0$ & $1.4$ -- $1.5$ & \\[0.2mm]
 & $\mathsf{T}_{156,23}$ & $0.37$ -- $0.40$ & $29$ -- $38$ & $29$ -- $39$ & $0.045$ -- $0.11$ & $0.52$ -- $0.58$ & $\checkmark$\\
 & & $0.68$ -- $0.71$ & $28$ -- $38$ & $28$ -- $39$ & $0.032$ -- $0.093$ & $1.5$ -- $1.6$ & \\[0.2mm]
 & $\mathsf{T}_{234,16}$ & $0.0$ -- $1.0$ & $22$ -- $51$ & $18$ -- $30$ & $0.072$ -- $1.9$ & $0.0$ -- $2.0$ & $\checkmark$\\[0.2mm]
 & $\mathsf{T}_{236,14}$ & $0.0$ -- $1.0$ & $22$ -- $48$ & $19$ -- $30$ & $0.14$ -- $2.0$ & $0.026$ -- $1.4$ & $\checkmark$\\[0.2mm]
 & $\mathsf{T}_{246,13}$ & $0.0$ -- $1.0$ & $25$ -- $40$ & $12$ -- $26$ & $0.56$ -- $1.4$ & $0.22$ -- $1.8$ & $\checkmark$\\[0.2mm]
 & $\mathsf{T}_{256,14}$ & $0.38$ -- $1.0$ & $44$ -- $51$ & $13$ -- $15$ & $0.87$ -- $1.1$ & $0.0$ -- $2.0$ & -\\[0.2mm]
 & $\mathsf{T}_{345,16}$ & $0.0$ -- $0.78$ & $44$ -- $54$ & $13$ -- $15$ & $0.86$ -- $1.1$ & $0.0$ -- $2.0$ & -\\[0.2mm]
 & $\mathsf{T}_{346,12}$ & $0.0$ -- $1.0$ & $25$ -- $40$ & $12$ -- $26$ & $0.55$ -- $1.4$ & $0.14$ -- $1.8$ & $\checkmark$\\[0.2mm]
 & $\mathsf{T}_{356,12}$ & $0.60$ -- $1.0$ & $18$ -- $21$ & $3.8$ -- $5.2$ & $0.86$ -- $1.1$ & $0.64$ -- $1.1$ & -\\[0.2mm]
\hline
IO & $\mathsf{T}_{123,45}$ & $0.0$ -- $1.0$ & $7.0$ -- $11$ & $27$ -- $29$ & $0.63$ -- $1.4$ & $0.0$ -- $2.0$ & $\checkmark$\\[0.2mm]
 & $\mathsf{T}_{123,56}$ & $0.0$ -- $1.0$ & $7.4$ -- $11$ & $27$ -- $30$ & $0.63$ -- $1.4$ & $0.0013$ -- $2.0$ & $\checkmark$\\[0.2mm]
 & $\mathsf{T}_{125,46}$ & $0.13$ -- $0.43$ & $33$ -- $50$ & $35$ -- $39$ & $1.2$ -- $1.4$ & $0.10$ -- $0.80$ & $\checkmark$\\
 & & $0.59$ -- $0.86$ & $24$ -- $29$ & $27$ -- $30$ & $0.69$ -- $0.82$ & $1.4$ -- $1.8$ & \\[0.2mm]
 & $\mathsf{T}_{126,45}$ & $0.0$ -- $0.69$ & $20$ -- $30$ & $30$ -- $35$ & $0.67$ -- $1.4$ & $0.93$ -- $2.0$ & -\\[0.2mm]
 & $\mathsf{T}_{134,56}$ & $0.30$ -- $0.43$ & $53$ -- $60$ & $47$ -- $52$ & $0.53$ -- $0.59$ & $0.0$ -- $2.0$ & -\\
 & & $0.82$ -- $0.95$ & $53$ -- $61$ & $47$ -- $52$ & $0.54$ -- $0.58$ & $0.83$ -- $0.90$ & \\[0.2mm]
 & $\mathsf{T}_{135,46}$ & $0.53$ -- $0.83$ & $33$ -- $49$ & $35$ -- $39$ & $0.61$ -- $0.75$ & $1.2$ -- $1.9$ & $\checkmark$\\[0.2mm]
 & $\mathsf{T}_{234,56}$ & $0.0$ -- $0.32$ & $8.1$ -- $21$ & $32$ -- $49$ & $0.44$ -- $1.7$ & $0.0068$ -- $2.0$ & $\checkmark$\\
 & & $0.66$ -- $1.0$ & $8.0$ -- $20$ & $32$ -- $49$ & $0.54$ -- $1.7$ & $0.69$ -- $1.3$ & \\[0.2mm]
 & $\mathsf{T}_{236,45}$ & $0.0$ -- $1.0$ & $7.1$ -- $18$ & $28$ -- $47$ & $0.30$ -- $1.6$ & $0.0$ -- $2.0$ & $\checkmark$\\[0.2mm]
 & $\mathsf{T}_{245,36}$ & $0.0$ -- $0.095$ & $9.2$ -- $16$ & $37$ -- $42$ & $0.44$ -- $1.6$ & $0.19$ -- $1.9$ & $\checkmark$\\
 & & $0.74$ -- $0.87$ & $9.5$ -- $16$ & $37$ -- $42$ & $1.5$ -- $1.6$ & $1.1$ -- $1.2$ & \\[0.2mm]
 & $\mathsf{T}_{246,13}$ & $0.12$ -- $0.48$ & $3.7$ -- $45$ & $30$ -- $37$ & $0.57$ -- $0.76$ & $0.40$ -- $0.65$ & $\checkmark$\\
 & & $0.75$ -- $1.0$ & $5.0$ -- $47$ & $30$ -- $37$ & $0.57$ -- $1.4$ & $0.25$ -- $1.6$ & \\[0.2mm]
 & $\mathsf{T}_{246,35}$ & $0.0$ -- $1.0$ & $8.3$ -- $18$ & $38$ -- $41$ & $0.42$ -- $1.6$ & $0.0$ -- $2.0$ & $\checkmark$\\[0.2mm]
 & $\mathsf{T}_{256,34}$ & $0.0$ -- $1.0$ & $2.9$ -- $6.4$ & $36$ -- $42$ & $0.41$ -- $1.6$ & $0.0$ -- $2.0$ & -\\[0.2mm]
 & $\mathsf{T}_{345,26}$ & $0.0$ -- $1.0$ & $2.9$ -- $6.4$ & $36$ -- $42$ & $0.38$ -- $1.6$ & $0.0$ -- $2.0$ & -\\[0.2mm]
 & $\mathsf{T}_{346,12}$ & $0.0$ -- $0.25$ & $12$ -- $50$ & $30$ -- $38$ & $0.60$ -- $0.77$ & $0.36$ -- $0.65$ & $\checkmark$\\
 & & $0.68$ -- $0.88$ & $11$ -- $51$ & $30$ -- $38$ & $1.2$ -- $1.4$ & $1.4$ -- $1.6$ & \\[0.2mm]
 & $\mathsf{T}_{346,25}$ & $0.0$ -- $1.0$ & $8.3$ -- $21$ & $38$ -- $42$ & $0.42$ -- $1.6$ & $0.0$ -- $2.0$ & -\\[0.2mm]
 & $\mathsf{T}_{356,24}$ & $0.14$ -- $0.31$ & $8.5$ -- $14$ & $37$ -- $42$ & $0.39$ -- $0.49$ & $0.80$ -- $0.98$ & $\checkmark$\\
 & & $0.85$ -- $1.0$ & $8.6$ -- $14$ & $38$ -- $42$ & $0.39$ -- $1.6$ & $0.13$ -- $1.8$ & \\[0.2mm]
 & $\mathsf{T}_{456,12}$ & $0.0$ -- $0.51$ & $1.5$ -- $2.7$ & $23$ -- $31$ & $0.60$ -- $1.4$ & $0.0$ -- $2.0$ & $\checkmark$\\
 & & $0.62$ -- $1.0$ & $1.5$ -- $2.7$ & $23$ -- $31$ & $0.69$ -- $1.4$ & $0.11$ -- $1.2$ & \\[0.2mm]
 & $\mathsf{T}_{456,13}$ & $0.026$ -- $1.0$ & $1.4$ -- $2.5$ & $24$ -- $33$ & $0.56$ -- $1.4$ & $0.0$ -- $2.0$ & $\checkmark$\\[0.2mm]
\hline
\end{tabular}}
\caption{\label{tab:T1D1S1} Neutrino mass textures in class $3_12_11_1$ compatible with data at $3\sigma$ CL.}
\end{table}

\begin{table}[h]
\centering
\scalebox{0.9}{
\begin{tabular}{|cccccccc|}
\hline
Spectrum & Texture & $\delta/\pi$ & $m$ [meV] & $m_{\beta\beta}$ [meV] & $\alpha_{21}/\pi$ & $\alpha_{31}/\pi$ & $1\sigma$ CL\\
\hline
NO & $\mathsf{T}_{1234}$ & $0.21$ -- $1.0$ & $20$ -- $51$ & $13$ -- $30$ & $0.54$ -- $1.4$ & $0.58$ -- $1.7$ & $\checkmark$\\[0.2mm]
& $\mathsf{T}_{1235}$ & $0.0$ -- $0.32$ & $30$ -- $34$ & $18$ -- $20$ & $0.64$ -- $1.4$ & $0.40$ -- $1.6$ & $\checkmark$\\
& & $0.63$ -- $0.96$ & $31$ -- $34$ & $18$ -- $20$ & $1.3$ -- $1.4$ & $1.4$ -- $1.6$ & \\[0.2mm]
& $\mathsf{T}_{1236}$ & $0.0$ -- $0.20$ & $20$ -- $44$ & $12$ -- $26$ & $0.56$ -- $1.5$ & $0.57$ -- $1.4$ & $\checkmark$\\
& & $0.57$ -- $0.79$ & $17$ -- $43$ & $11$ -- $26$ & $1.3$ -- $1.5$ & $0.36$ -- $0.53$ & \\[0.2mm]
& $\mathsf{T}_{1246}$ & $0.0$ -- $1.0$ & $26$ -- $41$ & $18$ -- $22$ & $0.58$ -- $1.4$ & $0.26$ -- $1.7$ & $\checkmark$\\[0.2mm]
& $\mathsf{T}_{1346}$ & $0.0$ -- $1.0$ & $26$ -- $41$ & $18$ -- $22$ & $0.60$ -- $1.4$ & $0.25$ -- $1.7$ & $\checkmark$\\[0.2mm]
\hline
IO & $\mathsf{T}_{2345}$ & $0.0$ -- $0.45$ & $11$ -- $13$ & $36$ -- $44$ & $0.37$ -- $1.6$ & $0.62$ -- $1.4$ & -\\
& & $0.73$ -- $1.0$ & $10$ -- $13$ & $36$ -- $44$ & $0.42$ -- $1.6$ & $0.063$ -- $1.8$ & \\[0.2mm]
& $\mathsf{T}_{2356}$ & $0.0$ -- $0.99$ & $11$ -- $13$ & $36$ -- $43$ & $0.38$ -- $1.6$ & $0.21$ -- $1.9$ & $\checkmark$\\[0.2mm]
& $\mathsf{T}_{2456}$ & $0.0$ -- $1.0$ & $1.1$ -- $1.9$ & $32$ -- $44$ & $0.39$ -- $1.7$ & $0.0$ -- $2.0$ & $\checkmark$\\[0.2mm]
& $\mathsf{T}_{3456}$ & $0.0$ -- $1.0$ & $1.1$ -- $1.8$ & $33$ -- $44$ & $0.32$ -- $1.6$ & $0.0$ -- $2.0$ & $\checkmark$\\[0.2mm]
\hline
\end{tabular}}
\caption{\label{tab:Q1S2} Neutrino mass textures in class $4_11_2$ compatible with data at $3\sigma$ CL.}
\end{table}


\begin{table}[h]
\centering
\scalebox{0.9}{
\begin{tabular}{|cccccccc|}
\hline
Spectrum & Texture & $\delta/\pi$ & $m$ [meV] & $m_{\beta\beta}$ [meV] & $\alpha_{21}/\pi$ & $\alpha_{31}/\pi$ & $1\sigma$ CL\\
\hline
NO & $\mathsf{T}_{12,35,46}$ & $0.0$ -- $1.0$ & $22$ -- $47$ & $8.4$ -- $31$ & $0.61$ -- $1.5$ & $0.42$ -- $1.7$ & $\checkmark$\\[0.2mm]
 & $\mathsf{T}_{13,25,46}$ & $0.0$ -- $0.99$ & $22$ -- $44$ & $8.5$ -- $29$ & $0.55$ -- $1.4$ & $0.38$ -- $1.6$ & $\checkmark$\\[0.2mm]
 & $\mathsf{T}_{14,23,56}$ & $0.33$ -- $0.51$ & $19$ -- $45$ & $20$ -- $46$ & $0.059$ -- $0.080$ & $0.52$ -- $0.66$ & $\checkmark$\\
 & & $0.68$ -- $0.72$ & $18$ -- $44$ & $20$ -- $45$ & $0.035$ -- $0.062$ & $1.4$ -- $1.6$ & \\[0.2mm]
 & $\mathsf{T}_{16,23,45}$ & $0.16$ -- $0.65$ & $12$ -- $39$ & $13$ -- $40$ & $1.9$ -- $2.0$ & $0.45$ -- $1.5$ & $\checkmark$\\[0.2mm]
\hline
IO & $\mathsf{T}_{14,23,56}$ & $0.32$ -- $1.0$ & $6.7$ -- $8.3$ & $12$ -- $24$ & $0.73$ -- $1.2$ & $0.27$ -- $1.1$ & -\\[0.2mm]
 & $\mathsf{T}_{16,23,45}$ & $0.0$ -- $0.68$ & $6.6$ -- $8.3$ & $12$ -- $28$ & $0.77$ -- $1.3$ & $0.87$ -- $1.7$ & $\checkmark$\\[0.2mm]
\hline
\end{tabular}}
\caption{\label{tab:D3} Neutrino mass textures in class $2_3$ compatible with data at $3\sigma$ CL.}
\end{table}

\begin{table}[h]
\centering
\scalebox{0.9}{
\begin{tabular}{|cccccccc|}
\hline
Spectrum & Texture & $\delta/\pi$ & $m$ [meV] & $m_{\beta\beta}$ [meV] & $\alpha_{21}/\pi$ & $\alpha_{31}/\pi$ & $1\sigma$ CL\\
\hline
NO & $\mathsf{T}_{123}$ & $0.0$ -- $1.0$ & $1.5$ -- $51$ & $4.8$ -- $30$ & $0.0$ -- $2.0$ & $0.0$ -- $2.0$ & $\checkmark$\\[0.2mm]
 & $\mathsf{T}_{124}$ & $0.0$ -- $1.0$ & $13$ -- $69$ & $11$ -- $36$ & $0.48$ -- $1.4$ & $0.25$ -- $1.7$ & $\checkmark$\\[0.2mm]
 & $\mathsf{T}_{125}$ & $0.0$ -- $0.99$ & $20$ -- $63$ & $16$ -- $39$ & $0.52$ -- $1.6$ & $0.0014$ -- $2.0$ & $\checkmark$\\[0.2mm]
 & $\mathsf{T}_{126}$ & $0.0$ -- $1.0$ & $13$ -- $62$ & $9.6$ -- $40$ & $0.36$ -- $1.6$ & $0.0$ -- $2.0$ & $\checkmark$\\[0.2mm]
 & $\mathsf{T}_{134}$ & $0.0$ -- $1.0$ & $17$ -- $67$ & $11$ -- $43$ & $0.47$ -- $1.5$ & $0.0$ -- $2.0$ & $\checkmark$\\[0.2mm]
 & $\mathsf{T}_{135}$ & $0.0$ -- $1.0$ & $20$ -- $64$ & $15$ -- $40$ & $0.38$ -- $1.5$ & $0.0$ -- $2.0$ & $\checkmark$\\[0.2mm]
 & $\mathsf{T}_{136}$ & $0.0$ -- $1.0$ & $11$ -- $67$ & $9.7$ -- $35$ & $0.57$ -- $1.5$ & $0.26$ -- $1.7$ & $\checkmark$\\[0.2mm]
 & $\mathsf{T}_{145}$ & $0.0$ -- $1.0$ & $28$ -- $63$ & $30$ -- $52$ & $0.0$ -- $2.0$ & $0.098$ -- $1.9$ & $\checkmark$\\[0.2mm]
 & $\mathsf{T}_{146}$ & $0.0$ -- $1.0$ & $15$ -- $70$ & $16$ -- $70$ & $0.0$ -- $2.0$ & $0.12$ -- $1.9$ & $\checkmark$\\[0.2mm]
 & $\mathsf{T}_{156}$ & $0.0$ -- $1.0$ & $27$ -- $63$ & $28$ -- $47$ & $0.0$ -- $2.0$ & $0.16$ -- $2.0$ & $\checkmark$\\[0.2mm]
 & $\mathsf{T}_{234}$ & $0.0$ -- $1.0$ & $18$ -- $71$ & $5.9$ -- $72$ & $0.0$ -- $2.0$ & $0.0011$ -- $2.0$ & $\checkmark$\\[0.2mm]
 & $\mathsf{T}_{235}$ & $0.0$ -- $1.0$ & $31$ -- $71$ & $9.5$ -- $71$ & $0.0$ -- $2.0$ & $0.0$ -- $2.0$ & $\checkmark$\\[0.2mm]
 & $\mathsf{T}_{236}$ & $0.0$ -- $1.0$ & $14$ -- $71$ & $4.6$ -- $71$ & $0.0$ -- $2.0$ & $0.0029$ -- $2.0$ & $\checkmark$\\[0.2mm]
 & $\mathsf{T}_{245}$ & $0.0$ -- $1.0$ & $20$ -- $66$ & $3.6$ -- $46$ & $0.59$ -- $1.5$ & $0.0017$ -- $2.0$ & $\checkmark$\\[0.2mm]
 & $\mathsf{T}_{246}$ & $0.0$ -- $1.0$ & $26$ -- $71$ & $7.1$ -- $71$ & $0.0$ -- $2.0$ & $0.0$ -- $2.0$ & $\checkmark$\\[0.2mm]
 & $\mathsf{T}_{256}$ & $0.0$ -- $1.0$ & $37$ -- $71$ & $7.4$ -- $33$ & $0.70$ -- $1.3$ & $0.0$ -- $2.0$ & $\checkmark$\\[0.2mm]
 & $\mathsf{T}_{345}$ & $0.0$ -- $1.0$ & $39$ -- $71$ & $8.5$ -- $36$ & $0.71$ -- $1.3$ & $0.0$ -- $2.0$ & $\checkmark$\\[0.2mm]
 & $\mathsf{T}_{346}$ & $0.0$ -- $1.0$ & $26$ -- $71$ & $7.1$ -- $71$ & $0.0$ -- $2.0$ & $0.0$ -- $2.0$ & $\checkmark$\\[0.2mm]
 & $\mathsf{T}_{356}$ & $0.012$ -- $1.0$ & $18$ -- $52$ & $3.5$ -- $37$ & $0.49$ -- $1.4$ & $0.0017$ -- $2.0$ & $\checkmark$\\[0.2mm]
\hline
IO & $\mathsf{T}_{123}$ & $0.0$ -- $1.0$ & $0.21$ -- $64$ & $27$ -- $47$ & $0.62$ -- $1.4$ & $0.0$ -- $2.0$ & $\checkmark$\\[0.2mm]
 & $\mathsf{T}_{125}$ & $0.0$ -- $1.0$ & $25$ -- $60$ & $27$ -- $47$ & $0.59$ -- $1.4$ & $0.030$ -- $2.0$ & $\checkmark$\\[0.2mm]
 & $\mathsf{T}_{126}$ & $0.0$ -- $1.0$ & $9.7$ -- $61$ & $27$ -- $48$ & $0.61$ -- $1.5$ & $0.028$ -- $2.0$ & $\checkmark$\\[0.2mm]
 & $\mathsf{T}_{134}$ & $0.0$ -- $1.0$ & $14$ -- $60$ & $29$ -- $47$ & $0.53$ -- $1.4$ & $0.025$ -- $1.9$ & $\checkmark$\\[0.2mm]
 & $\mathsf{T}_{135}$ & $0.0$ -- $1.0$ & $26$ -- $61$ & $29$ -- $48$ & $0.57$ -- $1.4$ & $0.11$ -- $2.0$ & $\checkmark$\\[0.2mm]
 & $\mathsf{T}_{145}$ & $0.0$ -- $1.0$ & $12$ -- $60$ & $11$ -- $40$ & $0.69$ -- $1.2$ & $0.0$ -- $2.0$ & $\checkmark$\\[0.2mm]
 & $\mathsf{T}_{146}$ & $0.0$ -- $0.99$ & $5.1$ -- $65$ & $12$ -- $67$ & $0.41$ -- $1.6$ & $0.020$ -- $2.0$ & $\checkmark$\\[0.2mm]
 & $\mathsf{T}_{156}$ & $0.0$ -- $1.0$ & $12$ -- $62$ & $11$ -- $38$ & $0.76$ -- $1.3$ & $0.0$ -- $2.0$ & $\checkmark$\\[0.2mm]
 & $\mathsf{T}_{234}$ & $0.0$ -- $1.0$ & $0.97$ -- $66$ & $30$ -- $82$ & $0.0$ -- $2.0$ & $0.0088$ -- $2.0$ & $\checkmark$\\[0.2mm]
 & $\mathsf{T}_{235}$ & $0.0$ -- $1.0$ & $3.5$ -- $65$ & $33$ -- $80$ & $0.0$ -- $2.0$ & $0.0$ -- $2.0$ & $\checkmark$\\[0.2mm]
 & $\mathsf{T}_{236}$ & $0.0$ -- $1.0$ & $0.66$ -- $66$ & $27$ -- $81$ & $0.0$ -- $2.0$ & $0.0028$ -- $2.0$ & $\checkmark$\\[0.2mm]
 & $\mathsf{T}_{245}$ & $0.0$ -- $1.0$ & $0.17$ -- $64$ & $28$ -- $64$ & $0.38$ -- $1.7$ & $0.0$ -- $2.0$ & $\checkmark$\\[0.2mm]
 & $\mathsf{T}_{246}$ & $0.0$ -- $1.0$ & $0.79$ -- $65$ & $27$ -- $81$ & $0.0015$ -- $2.0$ & $0.0076$ -- $2.0$ & $\checkmark$\\[0.2mm]
 & $\mathsf{T}_{256}$ & $0.0$ -- $1.0$ & $0.22$ -- $60$ & $32$ -- $55$ & $0.38$ -- $1.7$ & $0.0$ -- $2.0$ & $\checkmark$\\[0.2mm]
 & $\mathsf{T}_{345}$ & $0.0$ -- $1.0$ & $0.11$ -- $65$ & $34$ -- $54$ & $0.30$ -- $1.6$ & $0.0017$ -- $2.0$ & $\checkmark$\\[0.2mm]
 & $\mathsf{T}_{346}$ & $0.0$ -- $1.0$ & $0.14$ -- $66$ & $29$ -- $80$ & $0.0$ -- $2.0$ & $0.0$ -- $2.0$ & $\checkmark$\\[0.2mm]
 & $\mathsf{T}_{356}$ & $0.0$ -- $1.0$ & $0.34$ -- $62$ & $32$ -- $60$ & $0.28$ -- $1.6$ & $0.0040$ -- $2.0$ & $\checkmark$\\[0.2mm]
 & $\mathsf{T}_{456}$ & $0.0$ -- $1.0$ & $0.94$ -- $5.2$ & $12$ -- $50$ & $0.0$ -- $2.0$ & $0.0$ -- $2.0$ & $\checkmark$\\[0.2mm]
\hline
\end{tabular}}
\caption{\label{tab:T1S3} Neutrino mass textures in class $3_11_3$ compatible with data at $3\sigma$ CL.}
\end{table}

\clearpage
\section{Extra correlations among the elements of $\mathbf{m}_{\nu}$ in class $2_11_20_2$}
\label{sec:extracorr}

\begin{table}[b]
\centering
\begin{tabular}{|cccc|}
\hline
Texture & Correlation & $a,b,c\in \mathbb{C}$ & $a,b,c \in \mathbb{R}$\\
\hline
$\mathsf{T}^{1,2}_{45}\,(A_{1,45})$ & $c=2a$ or $c=3a$  & $\checkmark$ & -- \\[2mm]
$\mathsf{T}^{1,3}_{45}\,(A_{2,45})$ & $b=2a$ or $c=2a$ & $\checkmark$ & -- \\[2mm]
$\mathsf{T}^{1,2}_{46}\,(A_{1,46})$ & $c=2a$ & $\checkmark$ & $\checkmark$ \\[2mm]
$\mathsf{T}^{1,3}_{46}\,(A_{2,46})$ & $c=2a$ & $\checkmark$ & $\checkmark$ \\[2mm]
$\mathsf{T}^{1,2}_{56}\,(A_{1,56})$ & $b=2a$ or $c=2a$ & $\checkmark$ & -- \\[2mm]
$\mathsf{T}^{1,3}_{56}\,(A_{2,56})$ & $c=2a$ or $c=3a$ & $\checkmark$ & -- \\
\hline
\end{tabular}
\caption{\label{tab:extracorr} Particular cases of extra correlations that can be imposed on the textures of table~\ref{tab:D1S2Z2}.  In all cases, the resulting $\mathbf{m}_{\nu}$ contains only three physical parameters. The checkmark indicates whether the texture with the given correlation is compatible with neutrino oscillation data at $3\sigma$~CL.}
\end{table}

\begin{figure}[b]
	\begin{center}
	\subfigure[Texture~$\mathsf{T}^{1,2}_{45}$~$(A_{1,45})$]{\label{fig6}\includegraphics[width=0.48\textwidth]{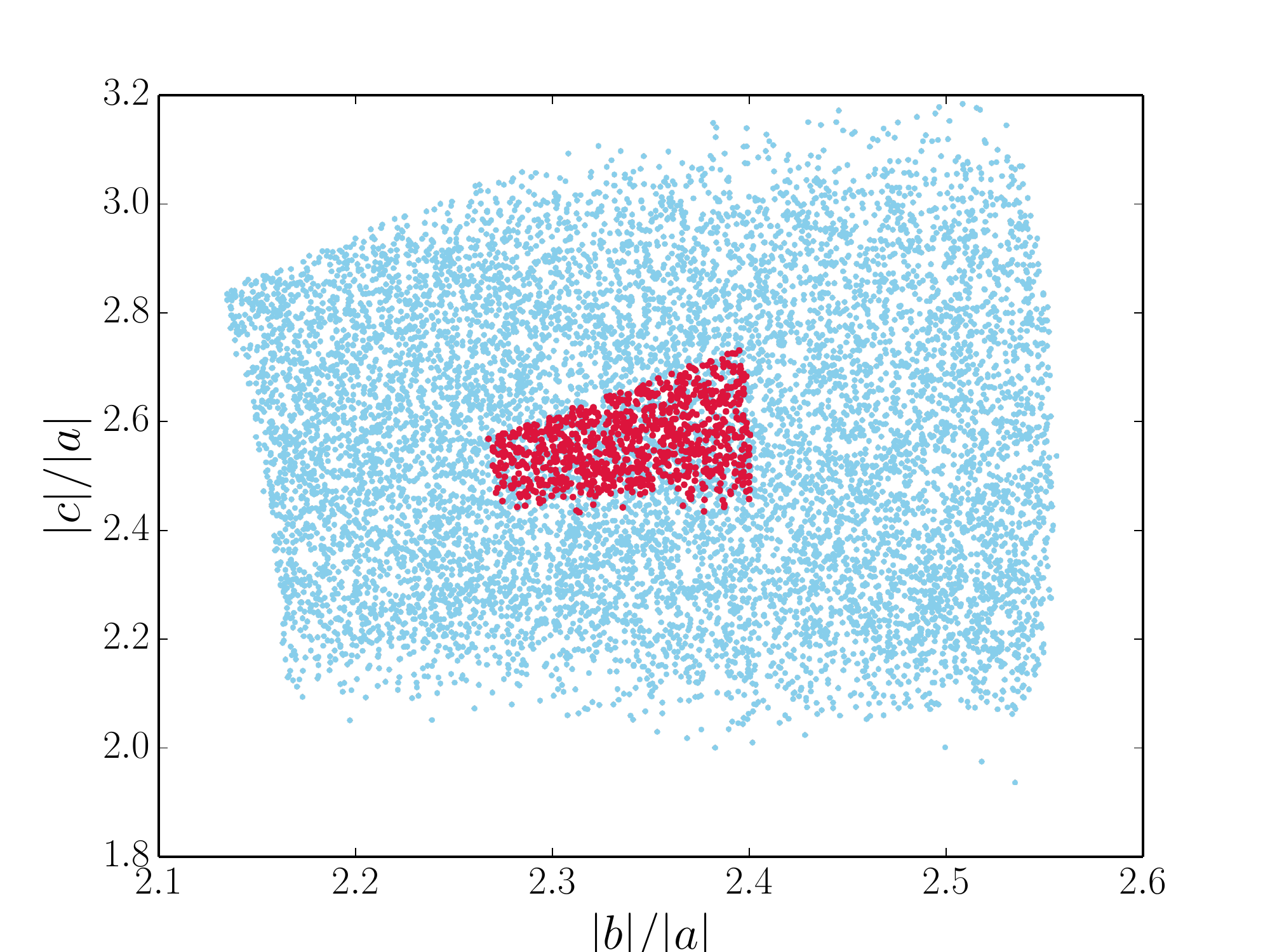}}
	\subfigure[Texture~$\mathsf{T}^{1,3}_{45}$~$(A_{2,45})$]{\label{fig7}\includegraphics[width=0.48\textwidth]{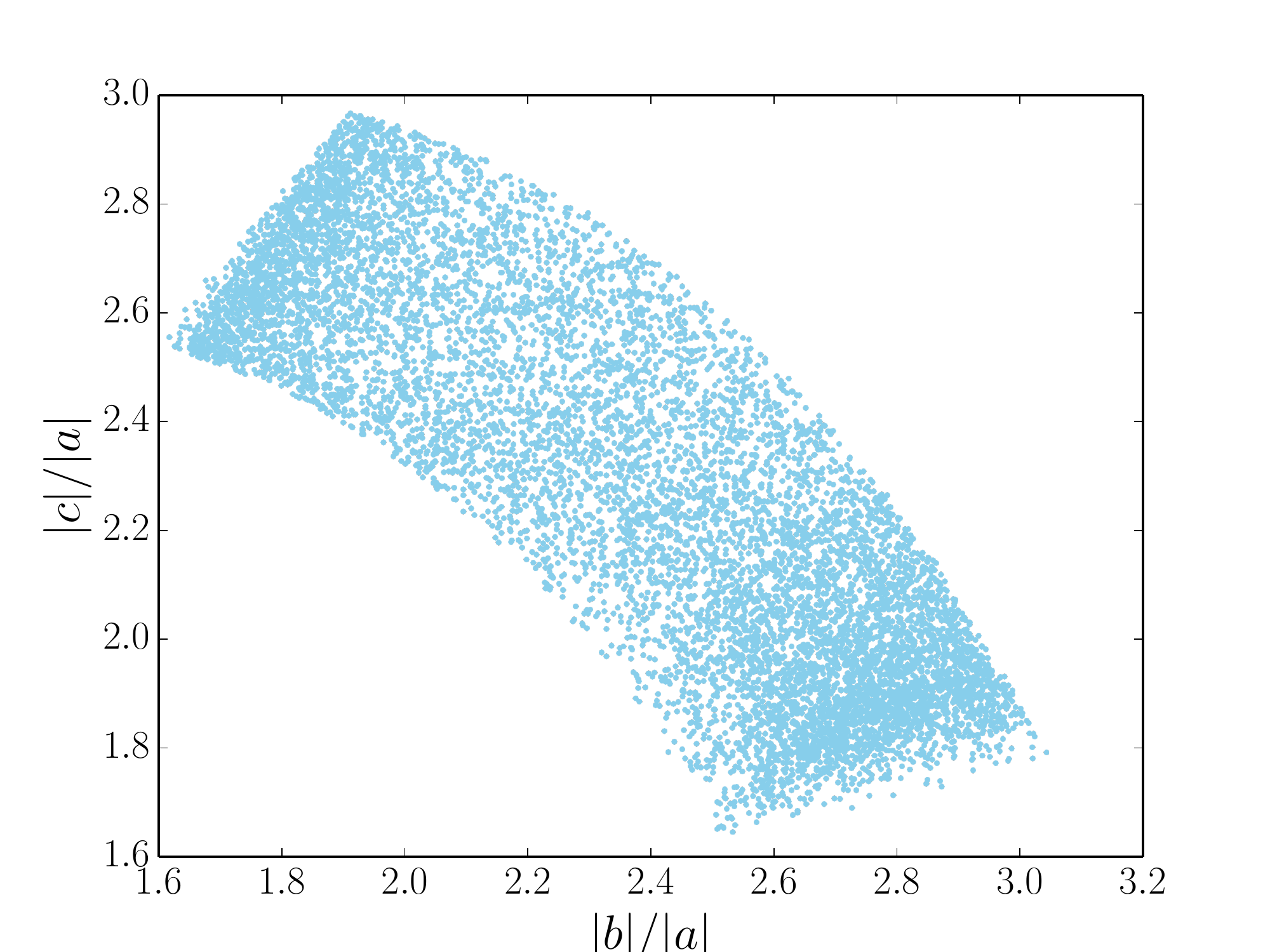}}
	
	\subfigure[Texture~$\mathsf{T}^{1,2}_{46}$~$(A_{1,46})$]{\label{fig8}\includegraphics[width=0.48\textwidth]{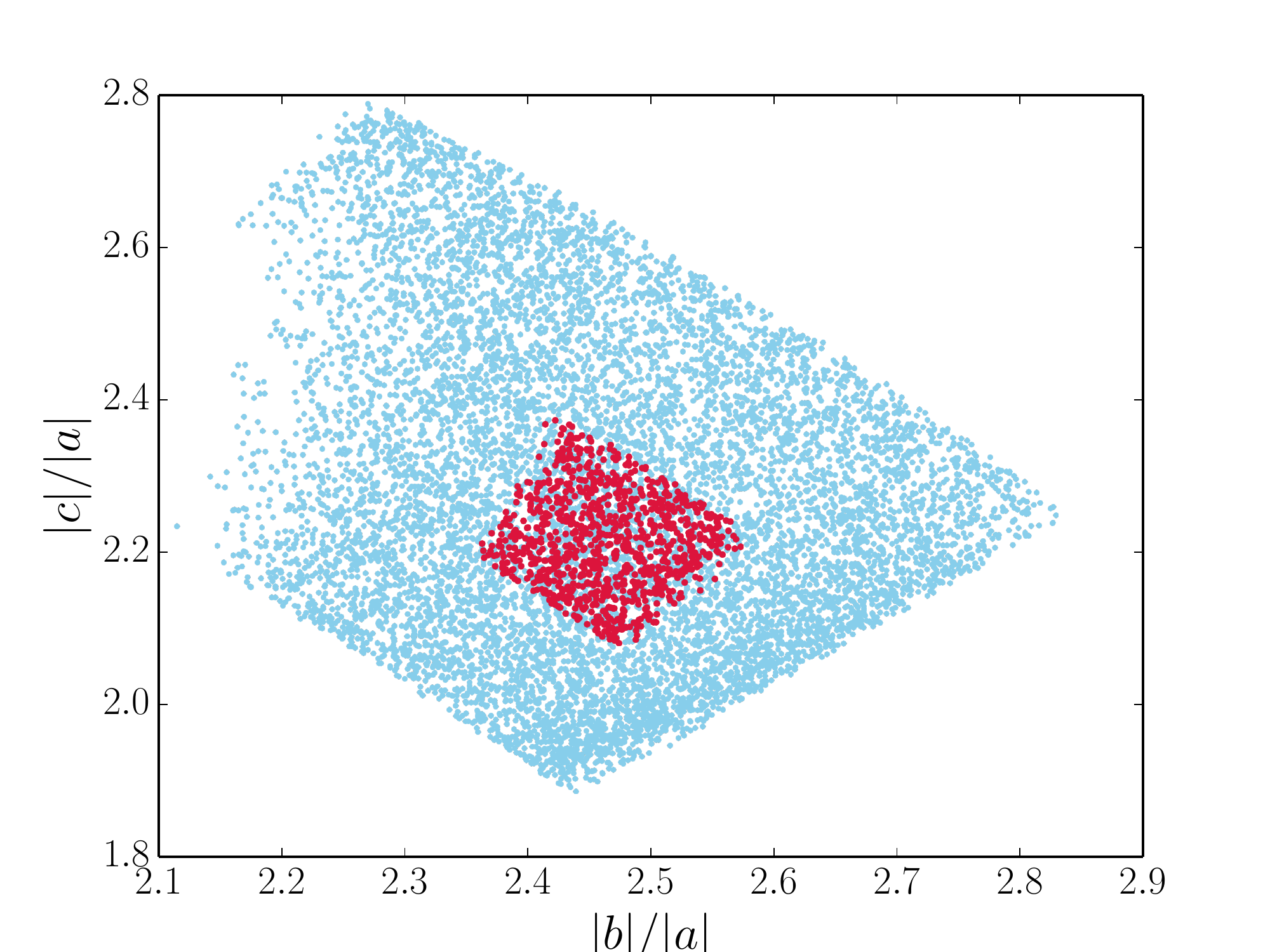}}
    \subfigure[Texture~$\mathsf{T}^{1,3}_{46}$~$(A_{2,46})$]{\label{fig9}\includegraphics[width=0.48\textwidth]{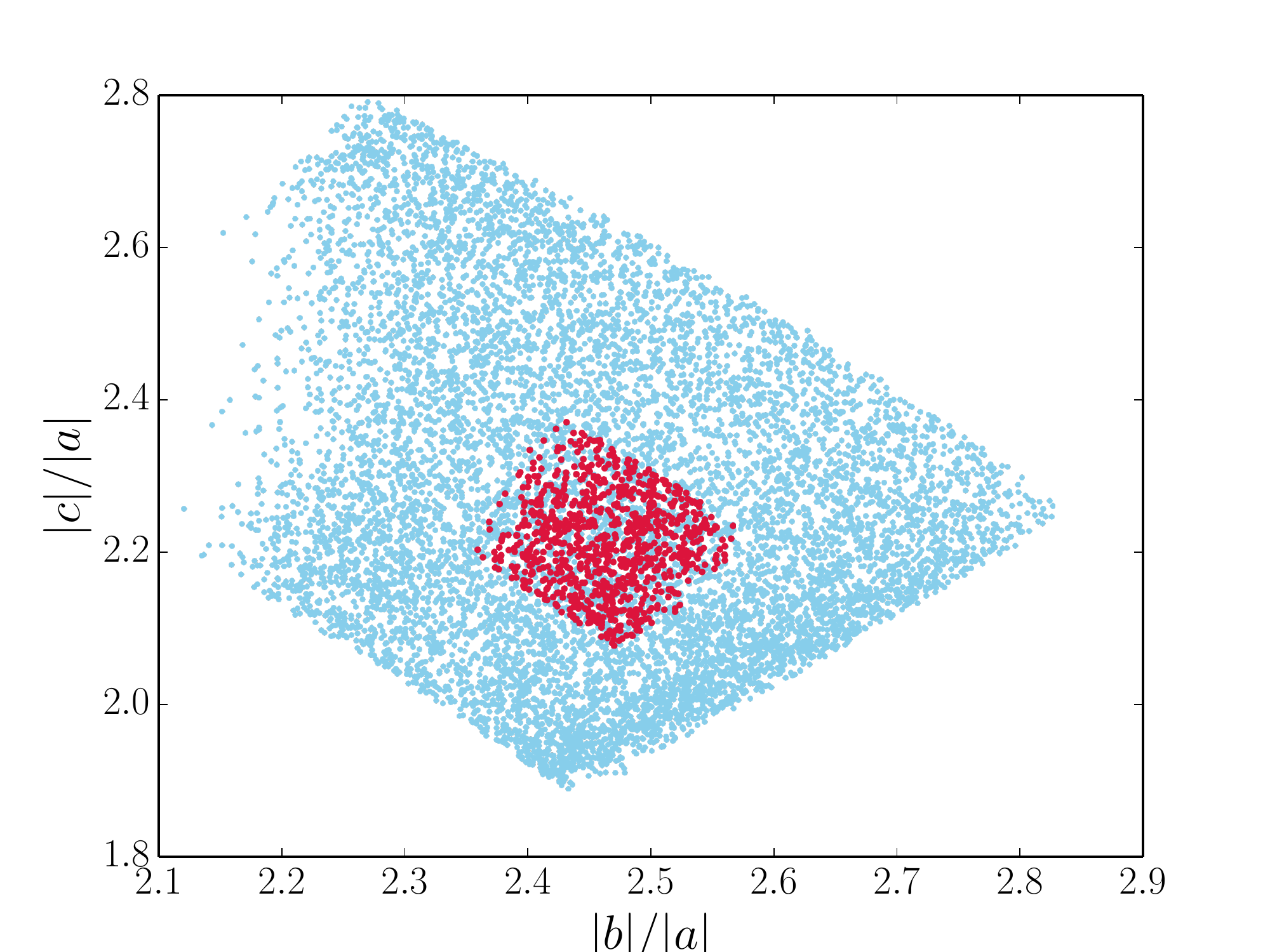}}
    
    \subfigure[Texture~$\mathsf{T}^{1,2}_{56}$~$(A_{1,56})$]{\label{fig10}\includegraphics[width=0.48\textwidth]{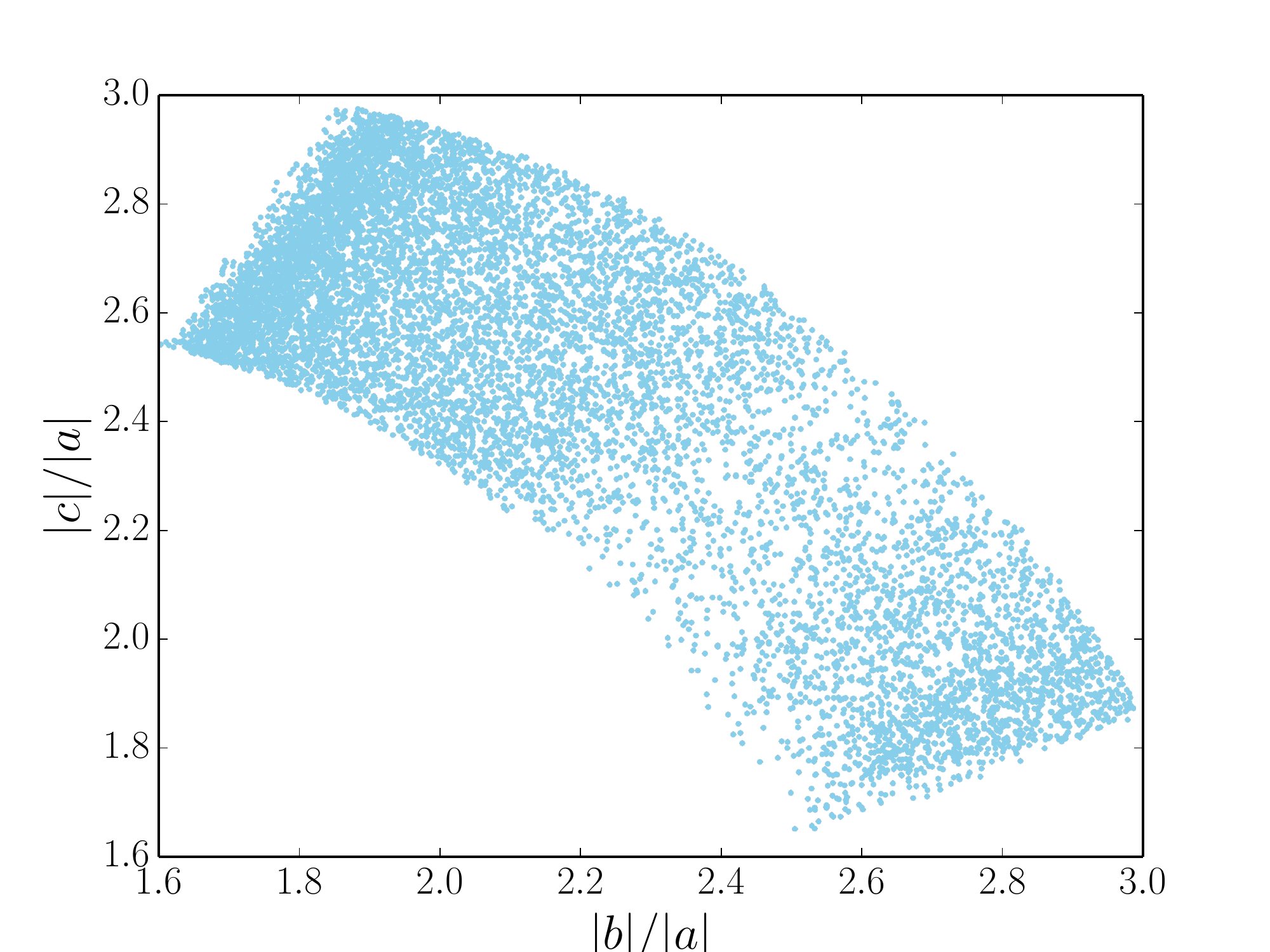}}
    \subfigure[Texture~$\mathsf{T}^{1,3}_{56}$~$(A_{2,56})$]{\label{fig11}\includegraphics[width=0.48\textwidth]{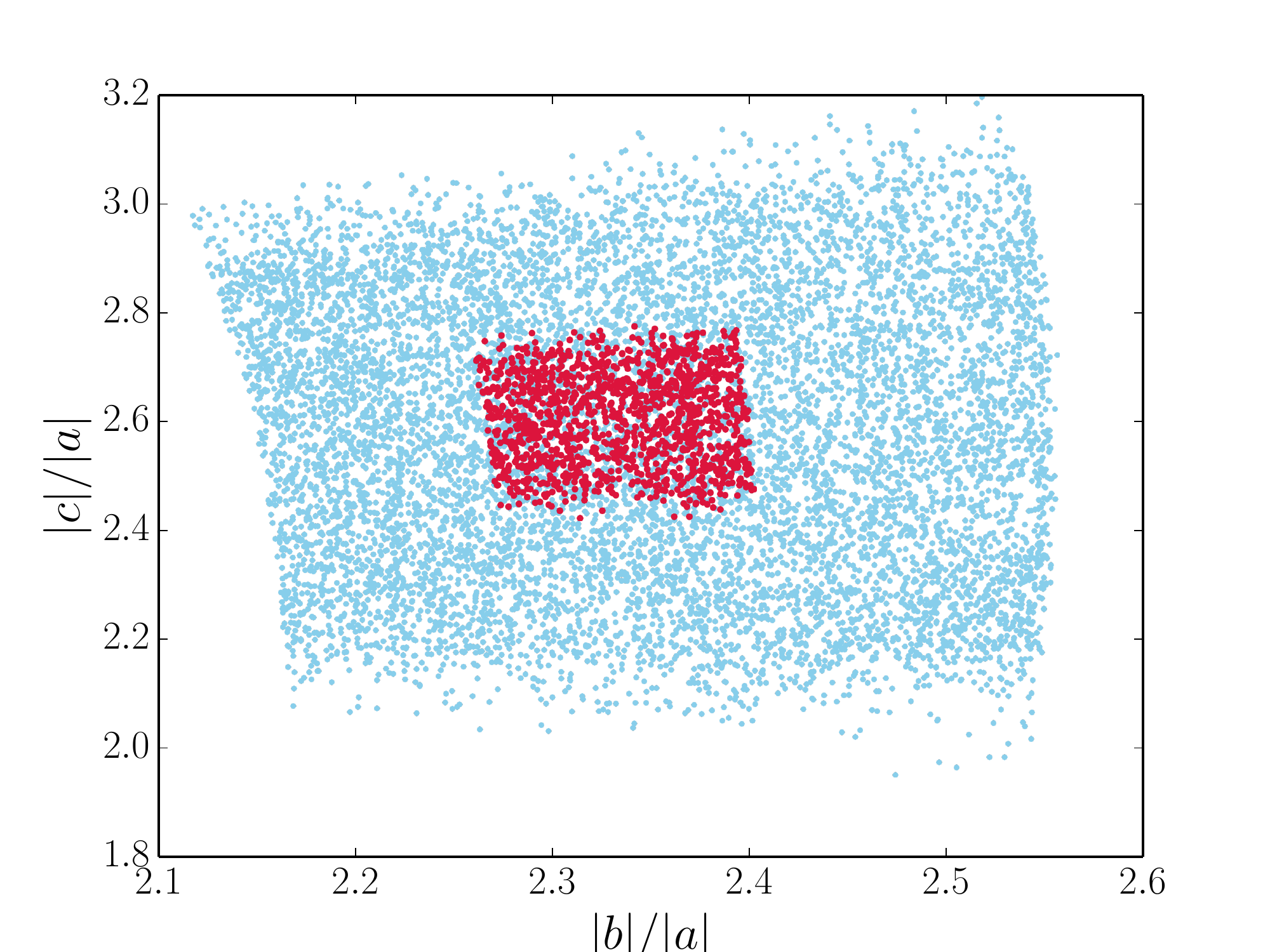}}

	\end{center}
	\caption{\label{fig3}Correlations among the moduli of the physical parameters of the viable textures within class~$2_11_20_2$ for NO spectrum (see table~\ref{tab:D1S2Z2}). The red and blue points in the scatter plots correspond to solutions compatible with neutrino data at the $1\sigma$ and $3\sigma$ CL, respectively.}
\end{figure}

In this appendix, we analyse the textures belonging to class~$2_11_20_2$ (cf. table~\ref{tab:D1S2Z2}) with the aim of looking for extra correlations among the matrix elements of $\mathbf{m}_{\nu}$ in the spirit of~\cite{Grimus:2012zm}. From our search, the textures
\begin{align}
\mathsf{T}^{1,2}_{45}&= \begin{pmatrix} 
0 & 0 & a \\ 
0 & b & b \\
a & b & c
\end{pmatrix}\,,
\quad  
\mathsf{T}^{1,3}_{45}=\begin{pmatrix} 
0 & a & 0 \\ 
a & b & b \\
0 & b & c
\end{pmatrix}\,,\\
\label{eq:A146}
\mathsf{T}^{1,2}_{46}&=\begin{pmatrix} 
0 & 0 & a \\ 
0 & b & c \\
a & c & b
\end{pmatrix}\,,
\quad  
\mathsf{T}^{1,3}_{46}=\begin{pmatrix} 
0 & a & 0 \\ 
a & b & c \\
0 & c & b
\end{pmatrix}\,,\\
\mathsf{T}^{1,2}_{56}&=\begin{pmatrix} 
0 & 0 & a \\ 
0 & c & b \\
a & b & b
\end{pmatrix}\,,\quad  
\mathsf{T}^{1,3}_{56}=\begin{pmatrix} 
0 & a & 0 \\ 
a & c & b \\
0 & b & b
\end{pmatrix}\,,
\end{align}
led to interesting correlations for NO neutrino mass spectrum. In figures~\ref{fig6}--\ref{fig11}, we plot the ratio $|c|/|a|$ versus the ratio $|b|/|a|$ for each of these textures. The plots show the allowed regions at the $1\sigma$ CL (red points) and $3\sigma$ CL (blue points). As can be seen from the figures, these ratios always lie approximately between $3/2$ and 3. Imposing an extra correlation among the input parameters $a$, $b$ and $c$, we have found several patterns that are compatible with data. These textures are summarised in table~\ref{tab:extracorr}. In addition to the two textures found in~\cite{Grimus:2012zm}, which correspond to the matrices~\eqref{eq:A146} with $c=2a$, there are eight new textures that are consistent with observations at $3\sigma$~CL, after reducing the number of physical parameters from four to three. Remarkably, the textures 
\begin{equation}
\begin{pmatrix} 
0 & 0 & 2a \\ 
0 & b & b \\
2a & b & 5a
\end{pmatrix}\,,\quad
\begin{pmatrix} 
0 & 0 & 2a \\ 
0 & 5a & b \\
2a & b & 5a
\end{pmatrix}\,,\quad
\begin{pmatrix} 
0 & 2a & 0 \\ 
2a & 5a & b \\
0 & b & 5a
\end{pmatrix}\,,\quad
\begin{pmatrix} 
0 & 2a & 0 \\ 
2a & 5a & b \\
0 & b & b
\end{pmatrix}\,,
\end{equation}
with $a$ real and  $b$ complex (or vice-versa), are compatible with observations even at $1\sigma$~CL.

\clearpage
\addcontentsline{toc}{section}{References}
\bibliographystyle{JHEP}
\bibliography{refs}

\end{document}